\documentstyle[preprint,aps]{revtex}
\begin{document}
\draft
\preprint{
\vbox{
\halign{&##\hfil\cr
	& AS-ITP-99-06 \cr
        & hep-ph/9901260 \cr
	& JAN. 1999 \cr}}
}
\title{A supersymmetric extension of the standard model with bilinear R-parity violation}
\author{Chao-hsi Chang$^{1,2}$, Tai-fu Feng$^{2}$\footnote{the contact author: fengtf@itp.ac.cn}}
\address{$^1$CCAST (World Laboratory), P.O.Box 8730,
Beijing 100080, P. R. China}
\address{$^2$Institute of Theoretical Physics, Academia Sinica, P.O.Box 2735,
Beijing 100080, P. R. China}
\maketitle
\begin{center}
\begin{abstract}
The minimum supersymmetric standard model with bilinear R-parity violation 
is studied systematically. Considering low-energy supersymmetry, we examine 
the structure of the bilinear R-parity violating model carefully. 
We analyze the mixing such as Higgs bosons
with sleptons, neutralinos with neutrinos and charginos with charged leptons in the
model. Possible and some important physics results such as the lightest Higgs 
may heavy than the weak Z-boson at tree level etc
are obtained. The Feynman rules for the model are derived in 
$'$t Hooft- Feynman gauge, which is convenient if perturbative calculations are needed beyond
the tree level.

\end{abstract}
\end{center}

\section{Introduction}
It is being increasingly realized by those engaged in the search for 
supersymmetry (SUSY)\cite{s1} that the principle of R-parity conservation, 
assumed to be sacrosanct in the prevalent search strategies, is not 
inviolable in practice. The R-parity of a particle is defined as $R=(-1)^
{2S+3B+L}$\cite{s2} and can be violated if either baryon (B) or lepton (L)
number is not conserved. In recent years, the intensive studies
of the supersymmetry that charactered by the bilinear R-parity violating
terms in the superpotential and the nonzero vacuum expectation values (VEVs)
of sneutrinos\cite{s3} have been undertaken. It stands as a simple 
supersymmetric (SUSY) model without R-parity which contains all
particles as those in the standard model, and can be arranged in a way that there is no
contradiction with the existing experimental data\cite{s4}. An impact of the R-parity
violation on the low energy phenomenology is twofold in the model. One leads the
lepton number violation (LNV) explicitly. The other is that the bilinear R-parity violation terms in
the superpotential and soft breaking terms generate nonzero vacuum
expectation values for the sneutrino fields $<\tilde{\nu_{i}}> \neq 0$ $(i=e,
\mu, \tau)$ and cause the new type mixing, such as neutrinos with neutralinos, 
charged leptons with charginos and sleptons with Higgs etc.

The R-conserving superpotential for the minimal supersymmetric standard 
model (MSSM) has the following form in superfields:
\begin{eqnarray}
{\cal W}_{MSSM} &=& \mu\varepsilon_{ij} \hat{H}_{i}^{1} \hat{H}_{j}^{2} + 
l_{I}\varepsilon_{ij} \hat{H}_{i}^{1} \hat{L}_{j}^{I} \hat{R}^{I} -
u_{I}(\hat{H}_{1}^{2} C^{JI*}\hat{Q}_{2}^{J} - \hat{H}_{2}^{2} \hat{Q}_{1}^{I}
)\hat{U}^{I}  \nonumber \\
 & & -d_{I}(\hat{H}_{1}^{1}\hat{Q}_{2}^{I} - \hat{H}_{2}^{1} C^{IJ} \hat{Q}_
{1}^{J})\hat{D}^{I}.
\label{eq-1}
\end{eqnarray}
Where $\hat{H}^{1}$, $\hat{H}^{2}$ are Higgs superfields; $\hat{Q}^{I}$
and $\hat{L}^{I}$ are quark and lepton superfields
respectively (I=1, 2, 3 is the index of generation), 
and all of them are in SU(2) weak-doublet. The rest superfields whereas 
$\hat{U}^{I}$ and $\hat{D}^{I}$ for quarks and $\hat{R}^{I}$ for charged leptons are in SU(2) weak-singlet. 
Here the indices i, j are contracted in a general way for SU(2) group and $C^{IJ}$ $(I, J=1, 2, 3)$ 
are the elements of the CKM matrix. However, when R-breaking interactions are considered, the superpotential
is modified as the follows\cite{s5}:
\begin{equation}
{\cal W} = {\cal W}_{MSSM} + {\cal W}_{L} + {\cal W}_{B}
\label{eq-2}
\end{equation}
with 
\begin{eqnarray}
&&{\cal W}_{L} = \varepsilon_{ij} [ \lambda_{IJK} \hat{L}_{i}^{I} \hat{L}_{j}^{J}
\hat{R}^{K} + \lambda_{IJK}^{\prime}\hat{L}_{i}^{I} \hat{Q}_{j}^{J} \hat{D}
^{K} + \epsilon_{I} \hat{H}_{i}^{2} \hat{L}_{j}^{I} ]  \nonumber \\
&&{\cal W}_{B} = \lambda_{IJK}^{\prime\prime} \hat{U}^{I} \hat{D}^{J} \hat{D}^
{K}.
\label{eq-3}
\end{eqnarray}

Since the proton decay experiments set down 
a very stringent limit on the byron number
violation\cite{s24}, 
we suppress the term ${\cal W}_{B}$ totally. The first two terms in
${\cal W}_{L}$ have received a lot of attention recently, and 
restrictions have been derived on them from existing experimental data\cite{s6}.
However, the term $\epsilon_{I} \varepsilon_{ij}\hat{H}_{i}^{2}\hat{L}_{j}^{J}$
is also a viable agent for R-parity breaking. It is particularly interesting 
because it can result in observable effects that are not to seen with
the trilinear terms alone. One of these distinctive effects is that, the 
lightest neutralino can decay invisibly into three neutrinos at the tree level, which is not
possible if only the trilinear terms in ${\cal W}_{L}$ are presented. The
significance of such bilinear R-parity violating interaction is further
emphasized by the following observations:
\begin{itemize}
\item Although it may seem possible
to rotate away the $\hat{H}_{2} \hat{L}$ terms by redefining the lepton and
Higgs superfields\cite{s7}, their effect is bound to show up via the soft
breaking terms.
\item Even if one may rotate these terms away at one energy scale, 
they will reappear at another one as the couplings evolving radiatively\cite{s8}. 
\item The
bilinear terms give rise to the trilinear terms at the one-loop level\cite{s9}.
\item It has been argued that if one wants to subsume R-parity violation in a
grand unified theory (GUT), then the trilinear R-parity violating terms come out
to be rather small in magnitude ($\sim 10^{-3}$ or so)\cite{s10}. However, the
superrenormalizable bilinear terms are not subjected to such requirements.
\end{itemize}

In this paper, we will keep $\epsilon_{I} \varepsilon_{ij}
\hat{L}_{i}^{I} H_{j}^{2}$ as the only R-parity violating terms 
to study the phenomenology of the model. The plan of this paper 
is follows. In Sect.II, we will describe
the basic ingredients of the supersymmetry with bilinear R-parity violation. The mass
matrices of the CP-even, CP-odd and charged Higgs are derived. Some interesting 
relations for CP-even and CP-odd Higgs masses are obtained.
For completeness, we also give the mixing matrices of charginos with charged leptons and
neutralinos with neutrinos. 
In Sect.III, we will give the Feynman rules for the interaction of the Higgs
bosons (sleptons) with the gauge bosons, and the charginos, neutralinos
with gauge bosons or Higgs bosons (sleptons). The self
interactions of the Higgs and the interactions of chargino (neutralino)-squark-quark 
are also given. In Sect.IV, we will analyze
the particle spectrum by the numerical method under a few assumptions about
the parameters in the model. We find that the possibility with large value for $
\epsilon_{3}$ and $\upsilon_{\tilde{\nu}_{\tau}}$ still survives under strong 
experimental restrictions for the masses of $\tau$-neutrino: 
$m_{\nu_{\tau}} \leq 20$ MeV and $\tau$-lepton: $m_{\tau}
= 1.77$ GeV. Finally we will close our discussions with comments on the model.

\section{The SUSY with bilinear R-parity violation}

As stated above, we are to consider a superpotential of the form:
\begin{eqnarray}
{\cal W} &=& \mu \varepsilon_{ij} \hat{H}_{i}^{1} \hat{H}_{j}^{2} + l_{I}
\varepsilon_{ij}\hat{H}_{i}^{1}\hat{L}_{j}^{I}\hat{R}^{I} - u^{I}(\hat{H}_{1}
^{2} C^{JI*}\hat{Q}_{2}^{J} - \hat{H}_{2}^{2}\hat{Q}_{1}^{I})\hat{U}^{I} \nonumber \\
 & & - d^{I}(\hat{H}_{1}^{1}\hat{Q}_{2}^{I} - \hat{H}_{2}^{1} C^{IJ} 
\hat{Q}_{1}^{J})\hat{D}^{I} + \epsilon_{I}\varepsilon_{ij}\hat{H}_{i}^{2}
\hat{L}_{j}^{I}
\label{eq-4}
\end{eqnarray}
with $\mu$, $\epsilon_{I}$ are the parameters with units of mass, $u^{I}$, $d^{I}$
and $l^{I}$ are the Yukawa couplings as in the MSSM with R-parity.
In order to break the supersymmetry, we introduce the soft SUSY-breaking terms:
\begin{eqnarray}
{\cal L}_{soft} & = & -m_{H^{1}}^{2}H_{i}^{1*}H_{i}^{1} - m_{H^{2}}^{2} H_{i}^{2*}
H_{i}^{2}-m_{L^{I}}^{2} \tilde{L}_{i}^{I*} \tilde{L}_{i}^{I} - m_{R^{I}}^{2}\tilde{R}^{I*} 
\tilde{R}^{I}  \nonumber  \\
 &  & -m_{Q^{I}}^{2} \tilde{Q}_{i}^{I*} \tilde{Q}_{i}^{I} - m_{D^{I}}^{2}
\tilde{D}
 ^{I*} \tilde{D}^{I} - m_{U^{I}}^{2}\tilde{U}^{I*} \tilde{U}^{I} + (m_{1}
\lambda_{B}
\lambda_{B}  \nonumber  \\
 &  & + m_{2}\lambda_{A}^{i}\lambda_{A}^{i} + m_{3} \lambda_{G}^{a}\lambda_{
 G}^{a} + h.c.) + \{ B\mu \varepsilon_{ij}H_{i}^{1}H_{j}^{2} + B_{I}\epsilon_{
I}\varepsilon_{ij}H_{i}^{2}\tilde{L}_{j}^{I}  \nonumber \\
 & & + \varepsilon_{ij} l_{sI}\mu H_{i}^{1}\tilde{L}_{j}^{I}\tilde{R}^{I} +
  d_{sI}\mu (-H_{1}^{1}\tilde{Q}_{2}^{I} + C^{IK}H_{2}^{1}\tilde{Q}_{1}^{K})
 \tilde{D}^{I}  \nonumber  \\
 & &+ u_{sI}\mu (-C^{KI*}H_{1}^{2}\tilde{Q}_{2}^{I} + H_{2}^{2}\tilde{Q}_{1}
 ^{I})
 \tilde{U}^{I} + h.c.\}
 \label{eq-5}
 \end{eqnarray}
where $m_{H^{1}}^{2}, m_{H^{2}}^{2}, m_{L^{I}}^{2}, m_{R^{I}}^{2}, m_{Q^{I}}^{2}, 
m_{D^{I}}^{2},$ and $m_{U^{I}}^{2}$ are the parameters with units of mass squared 
while $m_{3}, m_{2}, m_{1}$ denote the masses of $\lambda_{G}^{a}, \lambda_{A}^{i}$ and $\lambda_{B}$, 
the $SU(3)\times SU(2) \times U(1)$ gauginos. $B$ and $B_{I}$ are free parameters with 
units as mass. $d_{sI}, u_{sI}$, $l_{sI}$ $(I=1,2,3)$ 
are the soft breaking 
parameters that give the mass splitting between the quarks, leptons and their
supersymmetric partners. The rest parts (such as the part of gauge, 
matter and the gauge-matter interactions etc)
in the model are the same as the MSSM with R-parity 
and we will not repeat them here.

Thus the scalar potential of the model can be written as
\begin{eqnarray}
V &=& \sum_{i} |\frac{\partial{\cal W}}{\partial A_{i}}|^{2} + V_{D} + 
V_{soft}  \nonumber \\
 &=& V_{F} + V_{D} + V_{soft}
\label{eq-6}
\end{eqnarray}
where $A_{i}$ denote the scalar fields, $V_{D}$ is
the usual D-terms, $V_{soft}$ is the SUSY soft breaking terms given in 
Eq.\ (\ref{eq-5}). Using the superpotential Eq.\ (\ref{eq-4}) and the soft
breaking terms Eq.\ (\ref{eq-5}), we can write down the scalar potential precisely. 

The electroweak symmetry is broken spontaneously when the two Higgs doublets $H^{1}$, $H^{2}$ and the 
sleptons acquire nonzero vacuum expectation values (VEVs):
\begin{equation}
H^{1}=
\left( 
\begin{array}{c}
\frac{1}{\sqrt{2}}(\chi_{1}^{0} + \upsilon_{1} + i\phi_{1}^{0}) \\
H_{2}^{1}  \end{array}  \right)
\label{H1-vacuum}
\end{equation}
\begin{equation}
H^{2}=
\left(
\begin{array}{c}
H_{1}^{2}  \\
\frac{1}{\sqrt{2}}(\chi_{2}^{0} + \upsilon_{2} + i\phi_{2}^{0})  \end{array}  \right)
\label{H2-vacuum}
\end{equation}
and
\begin{equation}
\tilde{L}^{I} = 
\left(
\begin{array}{c}
\frac{1}{\sqrt{2}}(\chi_{\tilde{\nu}_{I}}^{0} + \upsilon_{\tilde{\nu}_{I}} + i\phi_{\tilde{\nu}_{I}}^{0})  \\
\tilde{L}_{2}^{I}    \end{array}  \right)
\label{L3-vacuum}
\end{equation}
where $\tilde{L}^{I}$ denote the slepton doublets and $I=e$, $\mu$, $\tau$, 
the generation indices of the leptons. 
From Eq.\ (\ref{eq-5}), Eq.\ (\ref{eq-6}), 
we can find the scalar potential includes the linear terms as following:
\begin{equation}
V_{tadpole} = t_{1}^{0}\chi_{1}^{0} + t_{2}^{0} \chi_{2}^{0} + t_{\tilde{\nu}_{e}}^{0} \chi_{\tilde{\nu}_{e}}^{0} + 
t_{\tilde{\nu}_{\mu}}^{0} \chi_{\tilde{\nu}_{\mu}}^{0} + t_{\tilde{\nu}_{\tau}}^{0} \chi_{\tilde{\nu}_{\tau}}^{0}
\label{tadpole0}
\end{equation}
where
\begin{eqnarray}
t_{1}^{0} &=& \frac{1}{8}(g^{2} + g^{\prime^{2}})\upsilon_{1}(\upsilon_{1}^{2} - \upsilon_{2}^{2} + 
\sum_{I}\upsilon_{\tilde{\nu}_{I}}^{2})
+ |\mu|^{2}\upsilon_{1} + m_{H^{1}}^{2}\upsilon_{1} - B\mu \upsilon_{2} -
\sum_{I} \mu \epsilon_{I}\upsilon_{I},  \nonumber \\
t_{2}^{0} &=& -\frac{1}{8}(g^{2} + g^{\prime^{2}})\upsilon_{2}(\upsilon_{1}^{2} - 
\upsilon_{2}^{2} + \sum_{I}\upsilon_{\tilde{\nu}_{I}}^{2})
+ |\mu|^{2}\upsilon_{2} + m_{H^{2}}^{2}\upsilon_{2} - B\mu \upsilon_{1} 
+ \sum_{I}\epsilon_{I}^{2}\upsilon_{2} \nonumber \\
 & & +\sum_{I} B_{I}\epsilon_{I}\upsilon_{I},     \nonumber \\
t_{\tilde{\nu}_{I}}^{0} &=& \frac{1}{8}(g^{2} + g^{\prime^{2}})
\upsilon_{\tilde{\nu}_{I}}(\upsilon_{1}^{2} - \upsilon_{2}^{2} + 
\sum_{I}\upsilon_{\tilde{\nu}_{I}}^{2})
+m_{L^{I}}^{2}\upsilon_{\tilde{\nu}_{I}} + \epsilon_{I}\sum_{J}
\epsilon_{J}\upsilon_{\tilde{\nu}_{J}}  \nonumber \\
 & & - \mu\epsilon_{I}
\upsilon_{1} + B_{I}\epsilon_{I}\upsilon_{2}.
\label{tadpole1}
\end{eqnarray}
Here $t_{i}^{0}$ ($i=1$, $2$, $\tilde{\nu}_{e}$, $\tilde{\nu}_{\mu}$, 
$\tilde{\nu}_{\tau}$) are tadpoles at the tree level and, 
the VEVs of the neutral scalar fields should satisfy the 
conditions $t_{i}^{0}=0$ ($i=1$, $2$, $\tilde{\nu}_{e}$, $\tilde{\nu}_{\mu}$, 
$\tilde{\nu}_{\tau}$), therefore one can obtain:
\begin{eqnarray}
m_{H^{1}}^{2} &=& -(|\mu|^{2} - \sum_{I}\epsilon_{I}\mu
\frac{\upsilon_{\tilde{\nu}_{I}}}{\upsilon_{1}} - B\mu\frac{\upsilon_{2}}{\upsilon_{1}}
 + \frac{1}{8}(g^{2} + g^{\prime^{2}})(\upsilon_{1}^{2} - 
 \upsilon_{2}^{2} + \sum_{I}\upsilon_{\tilde{\nu}_{I}}^{2})),  \nonumber \\
m_{H^{2}}^{2} &=& -(|\mu|^{2} + \sum_{I}\epsilon_{I}^{2} + 
\sum_{I}B_{I}\epsilon_{I}\frac{\upsilon_{\tilde{\nu}_{I}}}{\upsilon_{2}} - 
B\mu\frac{\upsilon_{1}}{\upsilon_{2}} - \frac{1}{8}(g^{2} 
+ g^{\prime^{2}})(\upsilon_{1}^{2} - \upsilon_{2}^{2} + 
\sum_{I}\upsilon_{\tilde{\nu}_{I}}^{2})) ,  \nonumber \\
m_{L^{I}}^{2} &=& -(\frac{1}{8}(g^{2} + g^{\prime^{2}})
(\upsilon_{1}^{2} - \upsilon_{2}^{2} - \sum_{I}\upsilon_{\tilde{\nu}_{I}}^{2})
+ \epsilon_{I}\sum_{J}\epsilon_{J}\frac{\upsilon_{\tilde{\nu}_{J}}}{\upsilon_{\tilde{\nu}_{I}}} - 
\epsilon_{I}\mu\frac{\upsilon_{1}}{\upsilon_{\tilde{\nu}_{I}}}  \nonumber \\
 & & +B_{I}\epsilon_{I}\frac{\upsilon_{2}}{
\upsilon_{\tilde{\nu}_{I}}}). \hspace{20mm} (I=e, \mu, \tau) \nonumber \\
\label{masspara}
\end{eqnarray}
For convenience, we will call all of these scalar bosons ($H^{1}$, $H^{2}$ 
and $\tilde{L}^{I}$) as Higgs below.
Now, we will give the Higgs boson mass matrix explicitly.
For the scalar sector, the mass squared matrices may be obtained by:
\begin{equation}
{\cal M}_{ij}^{2} = \frac{\partial^{2} V}{\partial \phi_{i} \phi_{j}}|_{minimum},
\label{getsmatrix}
\end{equation}
here "minimum" means to evaluate the values at $<H_{1}^{1}> = \frac{\upsilon_{1}}{\sqrt{2}}$, $<H_{2}^{2}> = 
\frac{\upsilon_{2}}{\sqrt{2}}$, 
$<\tilde{L}_{1}^{I}> = \frac{\upsilon_{\tilde{\nu}_{I}}}{\sqrt{2}}$ and $<A_{i}>=0$ ($A_{i}$ 
represent for all other scalar fields).
Thus the squared mass matrices of the CP-even and the CP-odd scalar bosons both 
are $5\times 5$, whereas matrix of
the charged Higgs is $8\times 8$.

\subsection{The squared mass matrices of Higgs}

From the scalar potential Eq.\ (\ref{eq-6}), we can find the mass terms:
\begin{equation}
{\cal L}_{m}^{even} = - \Phi_{even}^{\dag} {\cal M}_{even}^{2} \Phi_{even}
\label{l-add1}
\end{equation}
where "current" CP-even Higgs fields $\Phi_{even}^{T} = 
(\chi_{1}^{0}$, $\chi_{2}^{0}$, $\chi_{\tilde{\nu}_{e}}^{0}$, $\chi_{\tilde{\nu}_{\mu}}^{0}$, 
$\chi_{\tilde{\nu}_{\tau}}^{0})$. The mass matrix in Eq.\ (\ref{l-add1}) is 
\begin{equation}
{\cal M}_{even}^{2} = \left(
\begin{array}{ccccc}
r_{11} & -e_{12} - B\mu  &  e_{13} - \mu\epsilon_{1}  & e_{14} - \mu\epsilon_{2} & e_{15} - \mu\epsilon_{3} \\
-e_{12} - B\mu & r_{22} & -e_{23} + B_{1}\epsilon_{1} & -e_{24} + B_{2}\epsilon_{2} & -e_{25} + B_{3}\epsilon_{3} \\
e_{13} - \mu\epsilon_{1} & -e_{23} + B_{1}\epsilon_{1} & r_{33} & e_{34} + \epsilon_{1}\epsilon_{2} & e_{35} + \epsilon_{1}\epsilon_{3} \\
e_{14} - \mu\epsilon_{2} & -e_{24} + B_{2}\epsilon_{2} & e_{34} + \epsilon_{1}\epsilon_{2} & r_{44} & e_{45} + \epsilon_{2}\epsilon_{3} \\
e_{15} - \mu\epsilon_{3} & -e_{25} + B_{3}\epsilon_{3} & e_{35} + \epsilon_{1}\epsilon_{3} & e_{45} + \epsilon_{2}\epsilon_{3} & r_{55} 
\end{array}
\right)
\label{matrix-even}
\end{equation}
with notations
\begin{eqnarray}
r_{11} &=& \frac{g^{2} + g^{\prime^{2}}}{8}(3\upsilon_{1}^{2} - \upsilon_{2}^{2} + \sum_{I}\upsilon_{\tilde{\nu}_{I}}^{2}) +
|\mu|^{2} + m_{H^{1}}^{2}  \nonumber \\
 &=& \frac{g^{2} + g^{\prime^{2}}}{4}\upsilon_{1}^{2} + \sum_{I}\mu\epsilon_{I}\frac{\upsilon_{\tilde{\nu}_{I}}}{\upsilon_{1}}
+ B\mu\frac{\upsilon_{2}}{\upsilon_{1}},  \nonumber \\
r_{22} &=& \frac{g^{2} + g^{\prime^{2}}}{8}(-\upsilon_{1}^{2} + 3\upsilon_{2}^{2} - \sum_{I}\upsilon_{\tilde{\nu}_{I}}^{2}) +
|\mu|^{2} + \sum_{I}\epsilon_{I}^{2} + m_{H^{2}}^{2} \nonumber \\
 &=& \frac{g^{2} + g^{\prime^{2}}}{4}\upsilon_{2}^{2} + B\mu\frac{\upsilon_{1}}{\upsilon_{2}}
- \sum_{I}B_{I}\epsilon_{I}\frac{\upsilon_{\tilde{\nu}_{I}}}{\upsilon_{2}},  \nonumber \\
r_{33} &=& \frac{g^{2} + g^{\prime^{2}}}{8}(\upsilon_{1}^{2} - \upsilon_{2}^{2} + \sum_{I}\upsilon_{\tilde{\nu}_{I}}^{2} +
2 \upsilon_{\tilde{\nu}_{e}}^{2}) + \epsilon_{1}^{2} + m_{L^{1}}^{2}  \nonumber \\
 &=& \frac{g^{2} + g^{\prime^{2}}}{4}\upsilon_{\tilde{\nu}_{e}}^{2} + \mu\epsilon_{1}\frac{\upsilon_{1}}{\upsilon_{\tilde{\nu}_{e}}}
- B_{1}\epsilon_{1}\frac{\upsilon_{2}}{\upsilon_{\tilde{\nu}_{e}}} - \epsilon_{1}\epsilon_{2}\frac{\upsilon_{\tilde{\nu}_{\mu}}}{
\upsilon_{\tilde{\nu}_{e}}} - \epsilon_{1}\epsilon_{3}
\frac{\upsilon_{\tilde{\nu}_{\tau}}}{\upsilon_{\tilde{\nu}_{e}}}, \nonumber \\
r_{44} &=& \frac{g^{2} + g^{\prime^{2}}}{8}(\upsilon_{1}^{2} - \upsilon_{2}^{2} + \sum_{I}\upsilon_{\tilde{\nu}_{I}}^{2} +
2 \upsilon_{\tilde{\nu}_{\mu}}^{2}) + \epsilon_{2}^{2} + m_{L^{2}}^{2}  \nonumber \\
 &=& \frac{g^{2} + g^{\prime^{2}}}{4}\upsilon_{\tilde{\nu}_{\mu}}^{2} + \mu\epsilon_{2}\frac{\upsilon_{1}}{\upsilon_{\tilde{\nu}_{\mu}}}
- B_{2}\epsilon_{2}\frac{\upsilon_{2}}{\upsilon_{\tilde{\nu}_{\mu}}} - \epsilon_{1}\epsilon_{2}\frac{\upsilon_{\tilde{\nu}_{e}}}{
\upsilon_{\tilde{\nu}_{\mu}}} - \epsilon_{2}\epsilon_{3}
\frac{\upsilon_{\tilde{\nu}_{\tau}}}{\upsilon_{\tilde{\nu}_{\mu}}}, \nonumber \\
r_{55} &=& \frac{g^{2} + g^{\prime^{2}}}{8}(\upsilon_{1}^{2} - \upsilon_{2}^{2} + \sum_{I}\upsilon_{\tilde{\nu}_{I}}^{2} +
2 \upsilon_{\tilde{\nu}_{\tau}}^{2}) + \epsilon_{3}^{2} + m_{L^{3}}^{2}  \nonumber \\
 &=& \frac{g^{2} + g^{\prime^{2}}}{4}\upsilon_{\tilde{\nu}_{\tau}}^{2} + \mu\epsilon_{3}\frac{\upsilon_{1}}{\upsilon_{\tilde{\nu}_{\tau}}}
- B_{3}\epsilon_{3}\frac{\upsilon_{2}}{\upsilon_{\tilde{\nu}_{\tau}}} - \epsilon_{1}\epsilon_{3}\frac{\upsilon_{\tilde{\nu}_{e}}}{
\upsilon_{\tilde{\nu}_{\tau}}} - \epsilon_{2}\epsilon_{3}\frac{\upsilon_{\tilde{\nu}_{\mu}}}{\upsilon_{\tilde{\nu}_{\tau}}} 
\label{r12345}
\end{eqnarray}
and
\begin{eqnarray}
&&e_{12} = \frac{g^{2} + g^{\prime^{2}}}{4}\upsilon_{1}\upsilon_{2}, \hspace{10mm}
e_{13} = \frac{g^{2} + g^{\prime^{2}}}{4}\upsilon_{1}\upsilon_{\tilde{\nu}_{e}}, \nonumber \\
&&e_{14} = \frac{g^{2} + g^{\prime^{2}}}{4}\upsilon_{1}\upsilon_{\tilde{\nu}_{\mu}}, \hspace{10mm}
e_{15} = \frac{g^{2} + g^{\prime^{2}}}{4}\upsilon_{1}\upsilon_{\tilde{\nu}_{\tau}}, \nonumber \\
&&e_{23} = \frac{g^{2} + g^{\prime^{2}}}{4}\upsilon_{2}\upsilon_{\tilde{\nu}_{e}} ,\hspace{10mm}
e_{24} = \frac{g^{2} + g^{\prime^{2}}}{4}\upsilon_{2}\upsilon_{\tilde{\nu}_{\mu}} ,\nonumber \\
&&e_{25} = \frac{g^{2} + g^{\prime^{2}}}{4}\upsilon_{2}\upsilon_{\tilde{\nu}_{\tau}} ,\hspace{10mm}
e_{34} = \frac{g^{2} + g^{\prime^{2}}}{4}\upsilon_{\tilde{\nu}_{e}}\upsilon_{\tilde{\nu}_{\mu}} ,\nonumber \\
&&e_{35} = \frac{g^{2} + g^{\prime^{2}}}{4}\upsilon_{\tilde{\nu}_{e}}\upsilon_{\tilde{\nu}_{\tau}} ,\hspace{10mm}
e_{45} = \frac{g^{2} + g^{\prime^{2}}}{4}\upsilon_{\tilde{\nu}_{\mu}}\upsilon_{\tilde{\nu}_{\tau}} .\nonumber \\
\end{eqnarray}
In order to obtain the above mass matrix, the Eq.\ (\ref{masspara}) is used. 
The physical CP-even Higgs can be obtained by
\begin{equation}
H_{i}^{0} = \sum_{j=1}^{5}Z_{even}^{ij}\chi_{j}^{0}
\label{masseven}
\end{equation}
where $Z_{i,j}^{even}$ ($i$, $j=1$, $2$, $3$, $4$, $5$) are the elements of 
the matrix that converts the mass matrix Eq.\ (\ref{matrix-even}) into a diagonal one i.e.
translates the current fields into physical fields (corresponding to the eigenstates
of the mass matrix).

In the current basis $\Phi_{odd}^{T} = (\phi_{1}^{0}$, $\phi_{2}^{0}$, $\phi_{\tilde{\nu}_{e}}^{0}$,
$\phi_{\tilde{\nu}_{\mu}}^{0}$, $\phi_{\tilde{\nu}_{\tau}}^{0})$, 
the mass matrix for the CP-odd scalar fields can be written as:
\begin{equation}
{\cal M}_{odd}^{2} =
\left(
\begin{array}{ccccc}
s_{11} & B\mu & -\mu\epsilon_{1} & -\mu\epsilon_{2} & -\mu\epsilon_{3}  \\
B\mu & s_{22} & -B_{1}\epsilon_{1} & -B_{2}\epsilon_{2} & -B_{3}\epsilon_{3} \\
-\mu\epsilon_{1} & -B_{1}\epsilon_{1} & s_{33} & \epsilon_{1}\epsilon_{2} & \epsilon_{1}\epsilon_{3} \\
 -\mu\epsilon_{2} & -B_{2}\epsilon_{2} & \epsilon_{1}\epsilon_{2} & s_{44} & \epsilon_{2}\epsilon_{3}  \\
-\mu\epsilon_{3} & -B_{3}\epsilon_{3} & \epsilon_{1}\epsilon_{3} & \epsilon_{2}\epsilon_{3} & s_{55}
\end{array}
\right)  \label{massodd}
\end{equation}
with
\begin{eqnarray}
s_{11} &=& \frac{g^{2} + g^{\prime^{2}}}{8}(\upsilon_{1}^{2} - \upsilon_{2}^{2} + \sum_{I}\upsilon_{\tilde{\nu}_{I}}^{2}) + \mu^{2}
+ m_{H^{1}}^{2}  \nonumber \\
 &=& \sum_{I}\mu\epsilon_{I}\frac{\upsilon_{\tilde{\nu}_{I}}}{\upsilon_{1}} 
 + B\mu\frac{\upsilon_{2}}{\upsilon_{1}} ,\nonumber \\
s_{22} &=& -\frac{g^{2} + g^{\prime^{2}}}{8}(\upsilon_{1}^{2} - \upsilon_{2}^{2} + \upsilon_{\tilde{\nu}_{I}}^{2}) + \mu^{2} +
\sum_{I}\epsilon_{I}^{2} + m_{H^{2}}^{2}  \nonumber \\
 &=& B\mu\frac{\upsilon_{1}}{\upsilon_{2}} 
 - \sum_{I}B_{I}\epsilon_{I}\frac{\upsilon_{\tilde{\nu}_{I}}}{\upsilon_{2}} ,\nonumber \\
s_{33} &=& \frac{g^{2} + g^{\prime^{2}}}{8}(\upsilon_{1}^{2} - \upsilon_{2}^{2} + \sum_{I}\upsilon_{\tilde{\nu}_{I}}^{2}) + 
\epsilon_{1}^{2} + m_{L^{1}}^{2} \nonumber \\
 &=& \mu\epsilon_{1}\frac{\upsilon_{1}}{\upsilon_{\tilde{\nu}_{e}}} 
 - B_{1}\epsilon_{1}\frac{\upsilon_{2}}{\upsilon_{\tilde{\nu}_{e}}}
- \epsilon_{1}\epsilon_{2}\frac{\upsilon_{\tilde{\nu}_{\mu}}}{\upsilon_{\tilde{\nu}_{e}}} 
- \epsilon_{1}\epsilon_{3}\frac{\upsilon_{\tilde{\nu}_{\tau}}}{\upsilon_{\tilde{\nu}_{e}}}  ,\nonumber \\
s_{44} &=& \frac{g^{2} + g^{\prime^{2}}}{8}(\upsilon_{1}^{2} - \upsilon_{2}^{2} + \sum_{I}\upsilon_{\tilde{\nu}_{I}}^{2}) + 
\epsilon_{2}^{2} + m_{L^{2}}^{2} \nonumber \\
 &=& \mu\epsilon_{2}\frac{\upsilon_{1}}{\upsilon_{\tilde{\nu}_{\mu}}} - B_{2}\epsilon_{2}\frac{\upsilon_{2}}{\upsilon_{\tilde{\nu}_{\mu}}}
- \epsilon_{1}\epsilon_{2}\frac{\upsilon_{\tilde{\nu}_{e}}}{\upsilon_{\tilde{\nu}_{\mu}}} 
- \epsilon_{2}\epsilon_{3}\frac{\upsilon_{\tilde{\nu}_{\tau}}}{\upsilon_{\tilde{\nu}_{\mu}}}  ,\nonumber \\
s_{55} &=& \frac{g^{2} + g^{\prime^{2}}}{8}(\upsilon_{1}^{2} - \upsilon_{2}^{2} + \sum_{I}\upsilon_{\tilde{\nu}_{I}}^{2}) + 
\epsilon_{3}^{2} + m_{L^{3}}^{2} \nonumber \\
 &=& \mu\epsilon_{3}\frac{\upsilon_{1}}{\upsilon_{\tilde{\nu}_{\tau}}} - B_{3}\epsilon_{3}\frac{\upsilon_{2}}{\upsilon_{\tilde{\nu}_{\tau}}}
- \epsilon_{1}\epsilon_{3}\frac{\upsilon_{\tilde{\nu}_{e}}}{\upsilon_{\tilde{\nu}_{\tau}}} 
- \epsilon_{2}\epsilon_{3}\frac{\upsilon_{\tilde{\nu}_{\mu}}}{\upsilon_{\tilde{\nu}_{\tau}}}.  
\label{s12345}
\end{eqnarray}
From Eq.\ (\ref{massodd}) and Eq.\ (\ref{s12345}), one can find that the neutral Goldstone boson (with zero-mass) can be
given as\cite{s11}:
\begin{eqnarray}
G^{0} & \equiv & H_{6}^{0} \nonumber \\
&=& \sum_{i=1}^{5} Z_{odd}^{1,i} \phi_{i}^{0} \nonumber \\
  &=& \frac{1}{\upsilon}(\upsilon_{1}\phi_{1}^{0} - \upsilon_{2}\phi_{2}^{0} + 
  \upsilon_{\tilde{\nu}_{e}}\phi_{\tilde{\nu}_{e}}^{0}
 + \upsilon_{\tilde{\nu}_{\mu}}\phi_{\tilde{\nu}_{\mu}}^{0} + 
 \upsilon_{\tilde{\nu}_{\tau}}\phi_{\tilde{\nu}_{\tau}}^{0}),
\label{ngold}
\end{eqnarray}
which is indispensable for spontaneous breaking the EW gauge symmetry. 
Here the $\upsilon=\sqrt{\upsilon_{1}^{2} + \upsilon_{2}^{2} + 
\sum\limits_{I}\upsilon_{\tilde{\nu}_{I}}^{2}}$ and the mass of $Z$-boson
$M_{Z} = \frac{\sqrt{g^{2} + g^{\prime^{2}}}}{2}\upsilon$ is the same 
as R-parity conserved MSSM.  The other four physical neutral bosons can be 
written as:
\begin{equation}
H_{5+i}^{0}(i=2, 3, 4, 5)=\sum_{j=1}^{5}Z_{odd}^{i,j}\phi_{j}^{0}
\label{oddhiggs}
\end{equation}
where $Z_{i,j}^{odd}$ ($i$, $j=1$, $2$, $3$, $4$, $5)$ is 
the matrix that converts the current fields into
the physical eigenstates.

From the eigenvalue equations, one can find two independent relations:
\begin{eqnarray}
  \sum_{i=1}^{5}m_{H_{i}^{0}}^{2}  &=& \sum_{i=2}^{5}m_{H_{5+i}^{0}}^{2} + m_{Z}^{2} ,\nonumber \\
  \prod_{i=1}^{5}m_{H_{i}^{0}}^{2}  &=& \Bigg[\frac{\upsilon_{1}^{2} - \upsilon_{2}^{2} + 
\sum\limits_{I=1}^{3}\upsilon_{\tilde{\nu}_{I}}^{2}}{\upsilon^{2}}\Bigg]^{2} m_{Z}^{2} 
\prod_{i=2}^{5}m_{H_{5+i}^{0}}^{2} .
\label{massrelation}
\end{eqnarray}
If we introduce the following notations:
\begin{eqnarray}
 & & \upsilon_{1} = \upsilon\cos\beta\cos\theta_{\upsilon}  \nonumber \\
 & & \upsilon_{2} = \upsilon\sin\beta  \nonumber \\
 & & \sqrt{\sum\limits_{I=1}^{3}\upsilon_{\nu_{I}}^{2}} = \upsilon\cos\beta\sin\theta_{\upsilon}
\label{defineang}
\end{eqnarray}
the second relation of Eq.\ (\ref{massrelation}) can be written as:
\begin{equation}
\prod_{i=1}^{5}m_{H_{i}^{0}}^{2} = \cos^{2}2\beta m_{Z}^{2} \prod_{i=2}^{5}m_{H_{5+i}^{0}}^{2}
\label{massrelation1}
\end{equation}
The first relation of Eq.\ (\ref{massrelation}) is also 
obtained in Ref\cite{s12}, whereas
we consider the second relation of Eq.\ (\ref{massrelation}) is also important, 
the two equations are independent  
restrictions on the masses of neutral Higgs bosons. 
For instance, from Eq.\ (\ref{massrelation}) and Eq.\ (\ref{massrelation1}), we have 
a upper limit on the mass of the lightest Higgs at tree level in the model:
\begin{equation}
m_{H_{1}^{0}}^{2} \leq m_{H_{n}^{0}}^{2}\Bigg(\frac{m_{Z}^{2}\cos^{2}2\beta}
{m_{H_{n}^{0}}^{2}}\Bigg)^{\frac{1}{n-1}}
\frac{1-\frac{1}{n-1}\frac{m_{Z}^{2}}{m_{H_{n}^{0}}^{2}}}{1-\frac{1}{n-1}
\bigg(\frac{m_{Z}^{2}\cos^{2}2\beta}{m_{H_{n}^{0}}^{2}}\bigg)^{\frac{1}{n-1}}}
\label{bound-masshiggs}
\end{equation}
where $n \geq 2$ is the number of the CP-even Higgs, $m_{H_{1}^{0}}$ is the 
mass of the lightest one among them and $m_{H_{n}^{0}}$
is the heaviest one. Some points should be noted about Eq.\ (\ref{bound-masshiggs}):
\begin{itemize}
  \item    From Eq.\ (\ref{massrelation}) and Eq.\ (\ref{massrelation1}), we can find 
           $m_{H_{n}^{0}}^{2} \geq m_{z}^{2}$.
  \item    When $n=2$ or $m_{H_{1}^{0}}^{2} = \cdots = m_{H_{n}^{0}}^{2} = m_{H_{n+2}^{0}}^{2}=\cdots =
  m_{H_{n+n}^{0}}^{2}=m_{Z}^{2}$, $\cos^{2}2\beta=1$, "=" is established.
  \item   In the case of MSSM with R-parity (n=2), 
  $m_{H_{1}^{0}}=m_{Z}^{2}\cos^{2}2\beta\frac{1-\frac{m_{Z}^{2}}{m_{H_{2}^{0}}
  ^{2}}}{1-\frac{m_{Z}^{2}}{m_{H_{2}^{0}}^{2}}\cos^{2}2\beta} 
  \leq m_{Z}^{2}\cos^{2}2\beta $ is recovered.
\end{itemize}
So when $n > 2$, we cannot 
imposed a upper limit on the $m_{H_{1}^{0}}$ as that for 
the R-parity conserved MSSM at the tree level, namely, for the later it is just the case n=2\cite{s19}:
\begin{equation}
m_{H_{1}^{0}}^{2} \leq m_{Z}^{2}\cos^{2} 2\beta \leq m_{Z}^{2}. \label{mssmhig}
\end{equation}
Considering experimental data, one cannot rule out large $\epsilon_{I}$ (I=1, 2, 3)\cite{s21,sd12}.
Further more, even if $\epsilon_{I} \ll \mu$, we still have no reason to assume 
$B_{I}\epsilon_{I} \ll B\mu$ in general case. In the MSSM with R-parity, the radiative corrections
to mass of the lightest Higgs are large\cite{sg1}, when complete one-loop corrections 
and leading two-loop corrections of ${\cal O}(\alpha\alpha_{s})$ are included, the Ref\cite{sg2}
gives the limit on the lightest Higgs mass: $m_{H_{1}^{0}} \leq 132$GeV. 
In the MSSM without R-parity, there is not so stringent restriction on the lightest Higgs even
at the tree level.

With the "current" basis $\Phi_{c}=(H_{2}^{1*}$, $H_{1}^{2}$, $\tilde{L}_{2}^{1*}$,  
$\tilde{L}_{2}^{2*}$, $\tilde{L}_{2}^{3*}$, $\tilde{R}^{1}$, $\tilde{R}^{2}$, $\tilde{R}^{3})$
and Eq.\ (\ref{eq-6}),
we can find the following mass terms in Lagrangian:
\begin{equation}
{\cal L}_{m}^{C} = -\Phi_{c}^{\dag}{\cal M}_{c}^{2}\Phi_{c}
\label{eq-20}
\end{equation}
and ${\cal M}_{c}^{2}$ is given in Appendix.A.
Diagonalizing the mass matrix for the charged Higgs bosons, 
we obtain the zero mass Goldstone boson state:
\begin{eqnarray}
H_{1}^{+} &=& \sum_{i=1}^{8}Z_{c}^{1,i}\Phi_{c}^{i} \nonumber \\
 &=& \frac{1}{\upsilon}(\upsilon_{1}H_{2}^{1*} - \upsilon_{2}H_{1}^{2} + \upsilon_{\tilde{\nu}_{e}}\tilde{L}_{2}^{1*} + 
 \upsilon_{\tilde{\nu}_{\mu}}\tilde{L}_{2}^{2*} + \upsilon_{\tilde{\nu}_{\tau}}\tilde{L}_{2}^{3*}),
\label{cgold}
\end{eqnarray}
together with the charge conjugate state $H_{1}^{-}$, which are indispensable to break 
electroweak symmetry and give $W^{\pm}$ bosons  masses. 
With the transformation matrix $Z_{c}^{ij}$ (converts from the current fields into 
the physical eigenstates basis), the other seven physical eigenstates
$H_{i}^{+}$ $(i=2$, $3$, $4$, $5$, $6$, $7$, $8)$ can be expressed as:
\begin{equation}
H_{i}^{+} = \sum_{j=1}^{8}Z_{c}^{i,j}\Phi^{c}_{j} \hspace{0.2in}(i, j=1, \cdots, 8).
\label{charhiggs}
\end{equation}

\subsection{The mixing of neutralinos with neutrino:}

Due to the lepton number violation in the MSSM without R-parity, fresh and 
interesting mixing of neutralinos with neutrinos and charginos with charged leptons
may happen.
The piece of Lagrangian responsible for the mixing of neutralinos with neutrinos is:
\begin{eqnarray}
{\cal L}_{\kappa_{i}^{0}}^{mass} &=& \{ ig\frac{1}{\sqrt{2}}\tau_{ij}^{i}\lambda^{i}_{A}\psi_{j}A_{i}^{*} 
+ig^{\prime}\sqrt{2}Y_{i}\lambda_{B}\psi_{i}A_{i}^{*} -\frac{1}{2}\frac{\partial^{2} {\cal W}}{
\partial A_{i} \partial A_{j}}\psi_{i}\psi_{j} + h.c. \} 
\nonumber \\
 & & + m_{1}(\lambda_{B}\lambda_{B} + h.c.) + m_{2}(\lambda_{A}^{i}\lambda_{A}^{i} + h.c.)
\label{massneutra}
\end{eqnarray}
where ${\cal W}$ is given by Eq.\ (\ref{eq-4}). $\tau^{i}/2$ and $Y_{i}$ are the generators 
of the SU(2)$\times$U(1) gauge group and $\psi_{i}$, 
$A_{i}$ stand for generic two-component fermions and scalar fields. 
Writing down the Eq.\ (\ref{massneutra}) explicitly, we
obtain:
\begin{equation}
{\cal L}_{\chi_{i}^{0}}^{mass} = -\frac{1}{2}(\Phi^{0})^{T}{\cal M}_{N}\Phi^{0} + h.c.
\label{massneutra1}
\end{equation}
with the current basis $(\Phi^{0})^{T} = (-i\lambda_{B}$, $-i\lambda_{A}^{3}$, 
$\psi_{H^{1}}^{1}$, $\psi_{H^{2}}^{2}$, $\nu_{e_{L}}$, $\nu_{\mu_{L}}$,
$\nu_{\tau_{L}})$ and
\begin{equation}
{\cal M}_{N} = \left(
\begin{array}{ccccccc}
2m_{1} & 0 & -\frac{1}{2}g^{\prime}\upsilon_{1} & \frac{1}{2}g^{\prime}\upsilon_{2} & -\frac{1}{2}g^{\prime}\upsilon_{\tilde{\nu}_{e}} &
-\frac{1}{2}g^{\prime}\upsilon_{\tilde{\nu}_{\mu}} & -\frac{1}{2}g^{\prime}\upsilon_{\tilde{\nu}_{\tau}} \\
0 & 2m_{2} & \frac{1}{2}g\upsilon_{1} & -\frac{1}{2}g\upsilon_{2} & \frac{1}{2}g\upsilon_{\tilde{\nu}_{e}} 
& \frac{1}{2}g\upsilon_{\tilde{\nu}_{\mu}} & \frac{1}{2}g\upsilon_{\tilde{\nu}_{\tau}}  \\
-\frac{1}{2}g^{\prime}\upsilon_{1} & \frac{1}{2}g\upsilon_{1} & 0 & -\frac{1}{2}\mu & 0  & 0 & 0\\
\frac{1}{2}g^{\prime}\upsilon_{2} & -\frac{1}{2}g\upsilon_{2} & -\frac{1}{2}\mu & 0 & \frac{1}{2}\epsilon_{1} 
& \frac{1}{2}\epsilon_{2} & \frac{1}{2}\epsilon_{3}   \\
-\frac{1}{2}g^{\prime}\upsilon_{\tilde{\nu}_{e}} & \frac{1}{2}g\upsilon_{\tilde{\nu}_{e}} & 0 & \frac{1}{2}\epsilon_{1} & 0 & 0 & 0\\
-\frac{1}{2}g^{\prime}\upsilon_{\tilde{\nu}_{\mu}} & \frac{1}{2}g\upsilon_{\tilde{\nu}_{\mu}} & 0 & \frac{1}{2}\epsilon_{2} & 0 & 0 & 0\\
-\frac{1}{2}g^{\prime}\upsilon_{\tilde{\nu}_{\tau}} & \frac{1}{2}g\upsilon_{\tilde{\nu}_{\tau}} & 0 & \frac{1}{2}\epsilon_{3} & 0 & 0 & 0 
\end{array} \right)
\label{neutralino-matrix}
\end{equation}
The mixing has the formulation:
\begin{eqnarray}
&&-i\lambda_{B} = Z_{N}^{1i}\tilde{\chi}_{i}^{0}, \hspace{20mm}
-i\lambda_{A}^{3} = Z_{N}^{2i}\tilde{\chi}_{i}^{0}, \nonumber \\
&&\psi_{H_{1}^{1}} = Z_{N}^{3i}\tilde{\chi}_{i}^{0}, \hspace{20mm}
\psi_{H_{2}^{2}} = Z_{N}^{4i}\tilde{\chi}_{i}^{0}, \nonumber \\
&&\nu_{e_{L}} = Z_{N}^{5i}\tilde{\chi}_{i}^{0}, \hspace{20mm}
\nu_{\mu_{L}} = Z_{N}^{6i}\tilde{\chi}_{i}^{0}, \nonumber \\
&&\nu_{\tau_{L}} = Z_{N}^{7i}\tilde{\chi}_{i}^{0} 
\label{mixing-neutralino}
\end{eqnarray}
and transformation matrix $Z_{N}$ has the property
\begin{eqnarray}
Z_{N}^{T}{\cal M}_{N}Z_{N} &=& diag(m_{\tilde{\kappa}_{1}^{0}}, m_{\tilde{\kappa}_{2}^{0}}, 
m_{\tilde{\kappa}_{3}^{0}}, m_{\tilde{\kappa}_{4}^{0}}, m_{\nu_{e}},m_{\nu_{\mu}},m_{\nu_{\tau}}).
\label{define-ZN}
\end{eqnarray}
For convenience, we formulate all the neutral fermions into four component spinors as the follows:
\begin{equation}
\nu_{e}=
\left(  \begin{array}{c}
\tilde{\chi}_{5}^{0} \\
\bar{\tilde{\chi}}_{5}^{0}  \end{array}  
\right)  \label{define-eneutrino}
\end{equation}
\begin{equation}
\nu_{\mu}=
\left(  \begin{array}{c}
\tilde{\chi}_{6}^{0} \\
\bar{\tilde{\chi}}_{6}^{0}  \end{array}  
\right)  \label{define-muneutrino}
\end{equation}
\begin{equation}
\nu_{\tau}=
\left(  \begin{array}{c}
\tilde{\chi}_{7}^{0} \\
\bar{\tilde{\chi}}_{7}^{0}  \end{array}  
\right)  \label{define-tauneutrino}
\end{equation}
\begin{equation}
\kappa_{i}^{0}(i=1, 2, 3, 4)=
\left(  \begin{array}{c}
\tilde{\chi}_{i}^{0} \\
\bar{\tilde{\chi}}_{i}^{0}  \end{array}  
\right)  \label{define-neutralino}
\end{equation}
From Eq.\ (\ref{neutralino-matrix}), we find that only one type neutrinos 
obtains mass from the mixing\cite{se6}, we can assume it is the $\tau$-neutrino.
One of the stringent restrictions comes from the bound 
that the mass of $\tau$-neutrino should be less than
20 MeV\cite{s14}. For convenience, sometimes we will call the mixing of neutralinos and 
neutrinos as neutralinos shortly late on.

\subsection{The mixing of charginos with charged leptons}

Similar to the mixing of neutralinos and neutrinos, charginos mix with the charged leptons and form a set of
charged fermions: $e^{-}$, $\mu^{-}$, $\tau^{-}$, $\kappa_{1}^{\pm}$, $\kappa_{2}^{\pm}$. 
In current basis, $\Psi^{+T}
= (-i\lambda^{+}$, $\tilde{H}_{2}^{1}$, $e_{R}^{+}$, $\mu_{R}^{+}$, $\tau_{R}^{+})$ 
and $\Psi^{-T}=(-i\lambda^{-}$, $\tilde{H}_{1}^{2}$,
$e_{L}^{-}$, $\mu_{L}^{-}$, $\tau_{L}^{-})$, the charged fermion mass 
terms in the Lagrangian can be written as\cite{s15}
\begin{equation}
{\cal L}_{\chi_{i}^{\pm}}^{mass} = -\Psi^{-T}{\cal M}_{C} \Psi^{+} + h.c. 
\label{masscharg1}
\end{equation}
and the mass matrix:
\begin{equation}
{\cal M}_{C} = \left(
\begin{array}{ccccc}
2m_{2} & \frac{e\upsilon_{2}}{\sqrt{2}S_{W}} & 0  & 0 & 0 \\
\frac{e\upsilon_{1}}{\sqrt{2}S_{W}} & \mu  & \frac{l_{1}\upsilon_{\tilde{\nu}_{e}}}{\sqrt{2}} & 
\frac{l_{2}\upsilon_{\tilde{\nu}_{\mu}}}{\sqrt{2}} & \frac{l_{3}\upsilon_{\tilde{\nu}_{\tau}}}{\sqrt{2}} \\
\frac{e\upsilon_{\tilde{\nu}_{e}}}{\sqrt{2}S_{W}} & \epsilon_{1} & \frac{l_{1}\upsilon_{1}}{\sqrt{2}} & 0 & 0 \\
\frac{e\upsilon_{\tilde{\nu}_{\mu}}}{\sqrt{2}S_{W}} & \epsilon_{2} & 0 & \frac{l_{2}\upsilon_{1}}{\sqrt{2}} & 0 \\
\frac{e\upsilon_{\tilde{\nu}_{\tau}}}{\sqrt{2}S_{W}} & \epsilon_{3} & 0 & 0 & \frac{l_{3}\upsilon_{1}}{\sqrt{2}}  
\end{array}  \right) .
\label{chargino-matrix}
\end{equation}
Here $S_{W}=\sin\theta_{W}$ and $\lambda^{\pm} = \frac{\lambda_{A}^{1} \mp i\lambda_{A}^{2}}{\sqrt{2}}$.
Two mixing matrices $Z_{+}$, $Z_{-}$ can be obtained by diagonalizing the mass matrix ${\cal M}_{c}$ i.e.
the product $(Z_{+})^{T}{\cal M}_{C}Z_{-}$ is diagonal matrix:
\begin{equation}
(Z_{+})^{T}{\cal M}_{C}Z_{-}= \left(
\begin{array}{ccccc}
m_{\kappa_{1}^{-}} & 0 & 0 & 0 &0 \\
0 & m_{\kappa_{2}^{-}} & 0 & 0 & 0 \\
0 & 0 & m_{e} & 0 & 0 \\
0 & 0 & 0 & m_{\mu} & 0 \\
0 & 0 & 0 & 0 & m_{\tau}
\end{array}  \right)
\label{dig-chargino}
\end{equation}
If we denote the mass eigenstates with $\tilde{\chi}$:
\begin{eqnarray}
&&-i\lambda_{A}^{\pm} = Z_{\pm}^{1i}\tilde{\chi}_{i}^{\pm}, \hspace{20mm}
\psi_{H^{2}}^{1} = Z_{+}^{2i}\tilde{\chi}_{i}^{+}, \nonumber \\
&&\psi_{H^{1}}^{2} = Z_{-}^{2i}\tilde{\chi}_{i}^{-}, \hspace{20mm}
e_{L} = Z_{-}^{3i}\tilde{\chi}_{i}^{-}, \nonumber \\
&&e_{R} = Z_{+}^{3i}\tilde{\chi}_{i}^{+}, \hspace{20mm}
\mu_{L} = Z_{-}^{4i}\tilde{\chi}_{i}^{-}, \nonumber \\
&&\mu_{R} = Z_{+}^{4i}\tilde{\chi}_{i}^{+}, \hspace{20mm}
\tau_{L} = Z_{-}^{5i}\tilde{\chi}_{i}^{-}, \nonumber \\
&&\tau_{R} = Z_{+}^{5i}\tilde{\chi}_{i}^{+}.
\label{mixing-chargino}
\end{eqnarray}
The four-component fermions are defined as:
\begin{equation}
\kappa_{i}^{+}(i=1, 2, 3, 4, 5)=
\left(  \begin{array}{c}
\tilde{\chi}_{i}^{+} \\
\bar{\tilde{\chi}}_{i}^{-}  \end{array}  
\right)  \label{define-chargino}
\end{equation}
where $\kappa_{1}^{\pm}$, $\kappa_{2}^{\pm}$ are the usual charginos and 
$\kappa_{i}^{\pm}$ ($i=3$, $4$, $5$) correspond to $e$, $\mu$ and
$\tau$ leptons respectively. For convenience, sometimes we will 
call the mixing of charginos with charged leptons as charginos 
shortly late on.

From the above analyses, we have achieved the mass spectrum of the neutralino - neutrinos, chargino - charged leptons, 
neutral Higgs - sneutrinos and charged Higgs - charged sleptons. For the interaction vertices are also important, 
thus we will give 
the Feynman rules which are different from those of the MSSM with R-parity in next section.

\section{Feynman rules for the R-parity violating interaction}

We have discussed the mass spectrum of the MSSM with bilinear R-parity violation.
Now, we are discussing the Feynman rules for the model that are different 
from those in MSSM with R-parity. We are working in the t$^{\prime}$Hooft-
Feynman gauge\cite{s16} which has the gauge fixed terms as:
\begin{eqnarray}
{\cal L}_{GF} &=& -\frac{1}{2\xi}\Big(\partial^{\mu}A_{\mu}^{3} + \xi M_{Z}C_{W}H_{6}^{0}\Big)^{2} - \frac{1}{2\xi}\Big(\partial^{
\mu}B_{\mu} - \xi M_{Z}S_{W}H_{6}^{0}\Big)^{2} - \frac{1}{2\xi}\bigg(\partial^{\mu}A_{\mu}^{1}  \nonumber \\
 & &+\frac{i}{\sqrt{2}}\xi M_{W}\Big(
H_{1}^{+} - H_{1}^{-}\Big)\bigg)^{2} -  \frac{1}{2\xi}\bigg(\partial^{\mu}A_{\mu}^{2} + 
\frac{1}{\sqrt{2}}\xi M_{W}\Big(H_{1}^{+} + H_{1}^{-}\Big)\bigg)^{2}  \nonumber \\
 &=& \Bigg\{-\frac{1}{2\xi}(\partial^{\mu}Z_{\mu})^{2} - \frac{1}{2\xi}(\partial^{\mu}F_{\mu})^{2} - \frac{1}{\xi}
(\partial^{\mu}W_{\mu}^{+})(\partial^{\mu}W_{\mu}^{-}) \Bigg\} - \Bigg\{M_{Z}H_{6}^{0}\partial^{\mu}Z_{\mu}  \nonumber \\
 & & +iM_{W}\Big(H_{1}^{+}
\partial^{\mu}W_{\mu}^{-} - H_{1}^{-}\partial^{\mu}W_{\mu}^{+}\Big)\Bigg\} - 
\Bigg\{ \frac{1}{2}\xi M_{Z}^{2}H_{6}^{0^{2}} - \xi M_{W}^{2}H_{1}^{+}H_{1}^{-} \Bigg\},
\label{gauge-fixed}
\end{eqnarray}
where $C_{W}=\cos\theta_{W}$ and $H_{6}^{0}$, $H_{1}^{\pm}$ were given 
as Eq. (\ref{ngold}) and Eq. (\ref{cgold}). 
By inserting Eq.\ (\ref{gauge-fixed}) into interaction Lagrangian, one obtains the desired vertices for the Higgs bosons.
If CP is conserved i.e. we assume the relevant parameters are real, one finds (by analyzing the $H_{i}^{0}f\bar{f}$ couplings)
that $H_{1}^{0}$, $H_{2}^{0}$, $H_{3}^{0}$ $H_{4}^{0}$, $H_{5}^{0}$ are scalars and $H_{6}^{0}$, $H_{7}^{0}$, $H_{8}^{0}$, 
$H_{9}^{0}$, $H_{10}^{0}$ are pseudoscalar.

\subsection{Feynman rules for Higgs (slepton)- gauge boson interactions}

\begin{figure}
\setlength{\unitlength}{1mm}
\begin{picture}(140,180)(30,30)
\put(-10,-10){\includegraphics{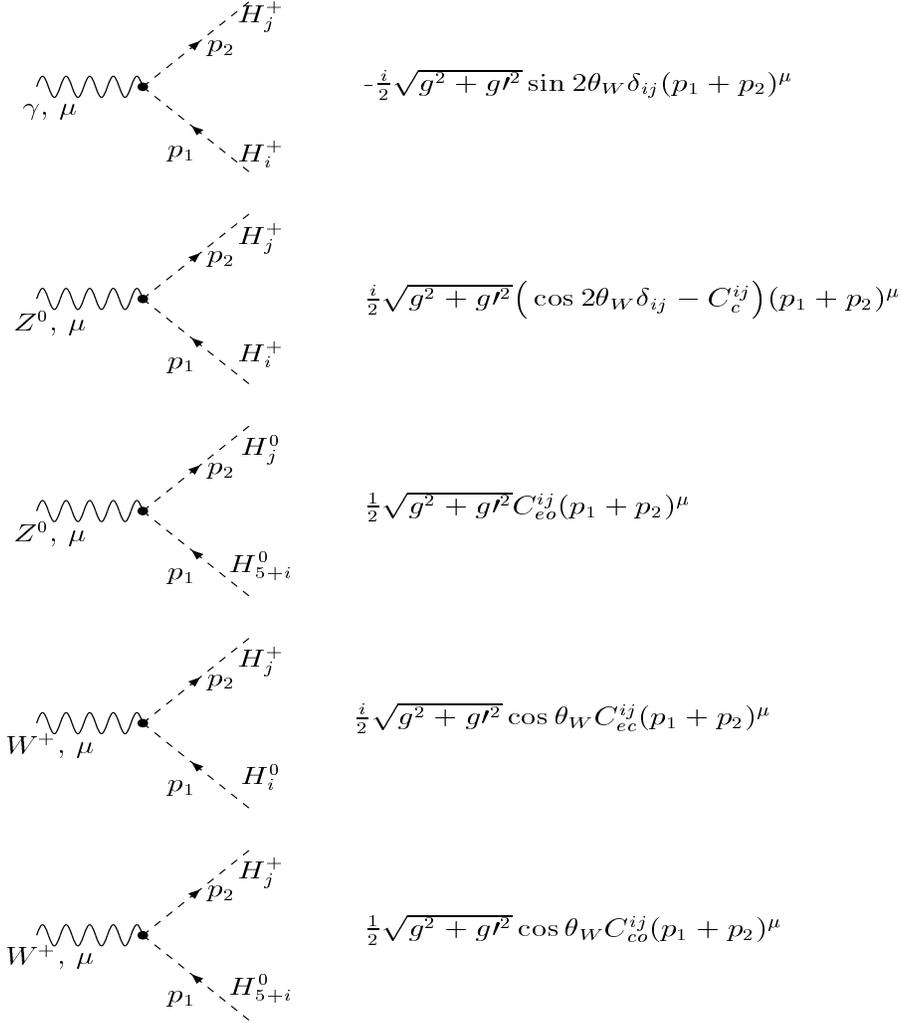}}
\end{picture}
\caption[]{Feynman rules for SSV vertices, the direction of momentum is indicated above}
\label{fig1}
\end{figure}

\begin{figure}
\setlength{\unitlength}{1mm}
\begin{picture}(140,180)(30,30)
\put(-10,-10){\includegraphics{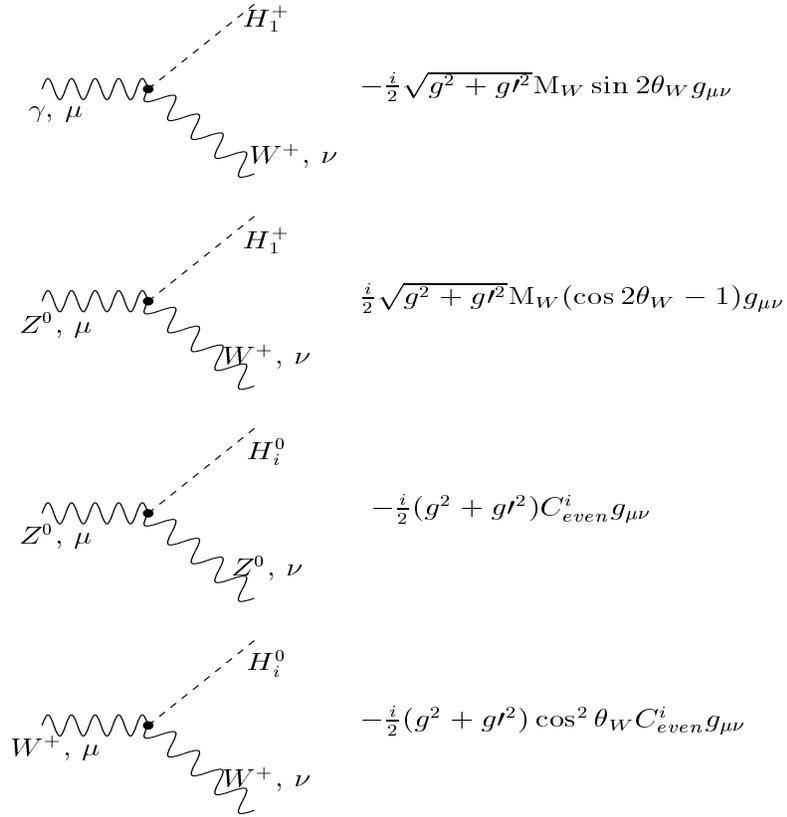}}
\end{picture}
\caption[]{Feynman rules for SVV vertices}
\label{fig2}
\end{figure}

Let us compute the vertices of Higgs (slepton)- gauge bosons in the model. 
The original interaction terms of Higgs bosons and
gauge bosons are given as
\begin{eqnarray}
{\cal L}_{int}^{1} &=& -\sum_{I}({\cal D}_{\mu}\tilde{L}^{I\dag}{\cal D}^{\mu}\tilde{L}^{I} - {\cal D}_{\mu}\tilde{R}^{I*}{\cal D}^{\mu}
\tilde{R}^{I}) - {\cal D}_{\mu}H^{1\dag}{\cal D}^{\mu}H^{1} - {\cal D}_{\mu}H^{2\dag}{\cal D}^{\mu}H^{2}  \nonumber \\
 &=&\sum_{I}\Bigg\{ \bigg[i\tilde{L}^{I\dag}\Big(g\frac{\tau^{i}}{2}A_{\mu}^{i} - 
\frac{1}{2}g^{\prime}B_{\mu}\Big)\partial^{\mu}\tilde{L}^{I} +
h.c. \bigg] - \tilde{L}^{I\dag}\Big(g\frac{\tau^{i}}{2}A_{\mu}^{i} \nonumber \\
&& - \frac{1}{2}g^{\prime}B_{\mu}\Big) \Big(g\frac{\tau^{j}}{2}A^{j\mu} - 
\frac{1}{2}g^{\prime}B^{\mu}\Big)\tilde{L}^{I} +  \Big( ig^{\prime} B_{\mu}\tilde{R}^{I*}\partial^{\mu}\tilde{R}^{I} \nonumber \\
&& +h.c. \Big) - g^{\prime^{2}}\tilde{R}^{I*}\tilde{R}^{I}B_{\mu}B^{\mu} \Bigg\} + 
\Bigg\{ H^{1\dag}\Big(g\frac{\tau^{i}}{2}A_{\mu}^{i} - \frac{1}{2}g^{\prime}B_{\mu}\Big)\partial^{\mu}H^{1} \nonumber \\
 & &+ h.c. \Bigg\}
- H^{1\dag}\Big(g\frac{\tau^{i}}{2}A_{\mu}^{i} - 
\frac{1}{2}g^{\prime}B_{\mu}\Big) \Big(g\frac{\tau^{j}}{2}A^{j\mu} -
\frac{1}{2}g^{\prime}B^{\mu}\Big)H^{1}  \nonumber \\
 & &+\Bigg\{ H^{2\dag}\Big(g\frac{\tau^{i}}{2}A_{\mu}^{i} - 
 \frac{1}{2}g^{\prime}B_{\mu}\Big)\partial^{\mu}H^{2} + h.c. \Bigg\}
- H^{2\dag}\Big(g\frac{\tau^{i}}{2}A_{\mu}^{i}  \nonumber \\
 & &- \frac{1}{2}g^{\prime}B_{\mu}\Big) \Big(g\frac{\tau^{j}}{2}A^{j\mu} -
\frac{1}{2}g^{\prime}B^{\mu}\Big)H^{2} \nonumber \\
&=& {\cal L}_{SSV} + {\cal L}_{SVV} + {\cal L}_{SSVV}.
\label{interaction1}
\end{eqnarray}
Here ${\cal L}_{SSV}$, ${\cal L}_{SVV}$ and ${\cal L}_{SSVV}$ represent the 
interactions in the physical basis, thus we have
\begin{eqnarray}
{\cal L}_{SSV} &=& \frac{i}{2}\sqrt{g^{2} + g^{\prime^{2}}}Z_{\mu} 
\Bigg\{ \partial^{\mu}\phi_{1}^{0}\chi_{1}^{0} - \phi_{1}^{0}
\partial^{\mu}\chi_{1}^{0} - \partial^{\mu}\phi_{2}^{0}\chi_{2}^{0} + 
\phi_{2}^{0}\partial^{\mu}\chi_{2}^{0} \nonumber  \\
 & &+\sum_{I}\Big(\partial^{\mu}\phi_{\tilde{\nu}_{I}}^{0}\chi_{\tilde{\nu}_{I}}^{0} - 
\phi_{\tilde{\nu}_{I}}^{0}\partial^{\mu}\chi_{\tilde{\nu}_{I}}^{0}\Big)\Bigg\} + \frac{1}{2}g\Bigg\{ 
W_{\mu}^{+}\bigg[\chi_{1}^{0}\partial^{\mu}H_{2}^{1} - \partial^{\mu}\chi_{1}^{0}H_{2}^{1} \nonumber \\
 & & -\chi_{2}^{0}\partial^{\mu}H_{1}^{2*} + \partial^{\mu}\chi_{2}^{0}H_{1}^{2*} + 
\sum_{I}\chi_{\tilde{\nu}_{I}}^{0}\partial^{\mu}\tilde{L}_{2}^{I} - 
\partial^{\mu}\chi_{\tilde{\nu}_{I}}^{0}\tilde{L}_{2}^{I} \bigg] + h.c.\Bigg\} \nonumber \\
 & & +\frac{i}{2}g\Bigg\{W_{\mu}^{+}
\bigg[\phi_{1}^{0}\partial^{\mu}H_{2}^{1*} - \partial^{\mu}\phi_{1}^{0}H_{2}^{1} + 
\phi_{2}^{0}\partial^{\mu}H_{1}^{2*} -   \partial^{\mu}\phi_{2}^{0}H_{1}^{2*} \nonumber \\
&&+ \sum_{I}(\phi_{\tilde{\nu}_{I}}^{0}\partial^{\mu}\tilde{L}_{2}^{I} -
\partial^{\mu}\phi_{\tilde{\nu}_{I}}^{0}\tilde{L}_{2}^{I}) \bigg] - h.c. \Bigg\} +  
\Bigg\{\frac{1}{2}\sqrt{g^{2} + g^{\prime^{2}}}\Big(\cos 2\theta_{W}Z_{\mu} \nonumber \\
 & &- \sin 2\theta_{W}A_{\mu}\Big)\bigg[\sum_{I}\Big(\tilde{L}_{2}^{I*}\partial^{\mu}
\tilde{L}_{2}^{I} -  \partial^{\mu}\tilde{L}_{2}^{I*}\tilde{L}_{2}^{I}\Big) - 
H_{1}^{2*}\partial^{\mu}H_{1}^{2}  \nonumber \\
 & &+\partial^{\mu}H_{1}^{2*}
H_{1}^{2} + H_{2}^{1*}\partial^{\mu}H_{2}^{1} - \partial^{\mu}H_{2}^{1*}H_{2}^{1}\bigg] + 
\Big(2\sin^{2}\theta_{W}Z_{\mu} \nonumber \\
 & &+ 2\sin\theta_{W}\cos\theta_{W}A_{\mu}\Big)\bigg[\sum_{I}\Big(\tilde{R}^{I*}\partial^{\mu}\tilde{R}^{I} - 
\partial^{\mu}\tilde{R}^{I*}\tilde{R}^{I}\Big)\bigg] \Bigg\} \nonumber \\
 &=& \frac{i}{2}\sqrt{g^{2} + g^{\prime^{2}}}C_{eo}^{ij}
\Big(\partial^{\mu}H_{5+i}^{0}H_{j}^{0} - H_{5+i}^{0}\partial^{\mu}H_{j}^{0}\Big)Z_{\mu}  \nonumber \\
 & &+\Bigg\{ \frac{1}{2}g C_{ec}^{ij}
\Big(H_{i}^{0}\partial^{\mu}H_{j}^{-} -\partial^{\mu}H_{i}^{0}H_{j}^{-}\Big)W_{\mu}^{+} + h.c. \Bigg\} \nonumber \\
 & & +\Bigg\{ \frac{i}{2}gC_{co}^{ij}
\Big(H_{5+i}^{0}\partial^{\mu}H_{j}^{-} - \partial^{\mu}H_{5+i}^{0}H_{j}^{-}\Big)W_{\mu}^{+}  \nonumber \\
 & &+ h.c. \Bigg\} + \Bigg\{\frac{1}{2}\sqrt{g^{2} + g^{\prime^{2}}}\bigg[\Big(
\cos 2\theta_{W}\delta^{ij}\nonumber \\
 & & -C_{c}^{ij}\Big)Z_{\mu}
\Big(H_{i}^{-}\partial^{\mu}H_{j}^{+} - \partial^{\mu}H_{i}^{-}H_{j}^{+}\Big) \nonumber \\
 & & - \sin 2\theta_{W}A_{\mu}\Big(H_{i}^{-}\partial^{\mu}H_{i}^{+} 
- \partial^{\mu}H_{i}^{-}H_{i}^{+}\Big)\bigg]\Bigg\},
\label{vertex-ssv}
\end{eqnarray}
with
\begin{eqnarray}
&&C_{eo}^{ij}= \sum_{\alpha=1}^{5}Z_{odd}^{i,\alpha}Z_{even}^{j,\alpha} 
- 2Z_{odd}^{i,2}Z_{even}^{j,2}, \nonumber \\
&&C_{ec}^{ij}=\sum_{\alpha=1}^{5}Z_{even}^{i,\alpha}Z_{c}^{j,\alpha} 
- 2Z_{even}^{i,2}Z_{c}^{j,2}, \nonumber \\
&&C_{co}^{ij}=\sum_{\alpha=1}^{5}Z_{odd}^{i,\alpha}Z_{c}^{j,\alpha} - 2Z_{odd}^{i,2}Z_{c}^{j,2}
, \nonumber \\
&&C_{c}^{ij}=\sum_{\alpha=6}^{8}Z_{c}^{i,\alpha}Z_{c}^{j,\alpha}.
\label{cdef}
\end{eqnarray}
Where the transformation matrices $Z_{even}$, $Z_{odd}$ and $Z_{c}$ are defined in section II. 
\begin{eqnarray}
{\cal L}_{SVV} &=& \frac{g^{2} + g^{\prime^{2}}}{4}\Big(\upsilon_{1}\chi_{1}^{0} + 
\upsilon_{2}\chi_{2}^{0}+\sum\limits_{I}\upsilon_{\tilde{\nu}_{I}}\chi_{\tilde{\nu}_{I}}^{0}\Big)
\Big(Z_{\mu}Z^{\mu} + 2\cos^{2}\theta_{W}W_{\mu}^{-}W_{+}^{\mu}\Big) \nonumber \\
 & &+\Bigg\{\frac{g^{2} + g^{\prime^{2}}}{4}\bigg[\cos\theta_{W}\Big(-1 + \cos 2\theta_{W}\Big)Z_{\mu}W_{+}^{\mu}
\Big(\upsilon_{1}H_{2}^{1} - \upsilon_{2}H_{1}^{2*} \nonumber \\
&&+ \sum_{I}\upsilon_{\tilde{\nu}_{I}}\tilde{L}_{2}^{I}\Big) - 
\cos\theta_{W}\sin 2\theta_{W}A_{\mu}W^{+\mu}\Big(\upsilon_{1}H_{2}^{1} - 
\upsilon_{2}H_{1}^{2*} \nonumber \\
 & &+ \sum_{I}\upsilon_{\tilde{\nu}_{I}}\tilde{L}_{2}^{I}\Big)\bigg] + h.c. \Bigg\} \nonumber \\
 &=& \frac{g^{2} + g^{\prime^{2}}}{4}C_{even}^{i}
\Big(H_{i}^{0}Z_{\mu}Z^{\mu} +2\cos^{2}\theta_{W}H_{i}^{0}W_{\mu}^{-}W^{+\mu}\Big) \nonumber \\
 & &-  \frac{g^{2} + g^{\prime^{2}}}{2}
S_{W}C_{W}\upsilon\bigg[S_{W}Z_{\mu}W^{+\mu}H_{1}^{-} + 
C_{W}A_{\mu}W^{+\mu}H_{1}^{-} + h.c. \bigg] ,
\label{vertex-svv}
\end{eqnarray}
with
\begin{eqnarray}
&&C_{even}^{i}=Z_{even}^{i,1}\upsilon_{1} + Z_{even}^{i,2}\upsilon_{2} + 
\sum_{I}Z_{even}^{i,I+2}\upsilon_{\tilde{\nu}_{I}} \label{ceven}
\end{eqnarray}
The piece of ${\cal L}_{SSVV}$ is given as
\begin{eqnarray}
{\cal L}_{SSVV} &=& -\frac{g^{2} + g^{\prime^{2}}}{4}
\bigg[\frac{1}{2}\Big(\chi_{1}^{0}\chi_{1}^{0} + \chi_{2}^{0}\chi_{2}^{0} + 
\sum_{I}\chi_{\tilde{\nu}_{I}}^{0}\chi_{\tilde{\nu}_{I}}^{0}\Big)Z_{\mu}Z^{\mu} \nonumber \\
 & &+ \cos^{2}\theta_{W}\Big(\chi_{1}^{0}\chi_{1}^{0} + \chi_{2}^{0}\chi_{2}^{0} + 
\sum_{I}\chi_{\tilde{\nu}_{I}}^{0}\chi_{\tilde{\nu}_{I}}^{0}\Big)W_{\mu}^{-}W^{+\mu} \bigg] - 
\frac{g^{2} + g^{\prime^{2}}}{4}\bigg[\frac{1}{2}\Big(\phi_{1}^{0}\phi_{1}^{0} \nonumber \\
 & &+ \phi_{2}^{0}\phi_{2}^{0} + 
\sum_{I}\phi_{\tilde{\nu}_{I}}^{0}\phi_{\tilde{\nu}_{I}}^{0}\Big)Z_{\mu}Z^{\mu} + 
\cos^{2}\theta_{W}\Big(\phi_{1}^{0}\phi_{1}^{0} + \phi_{2}^{0}\phi_{2}^{0} + 
\phi_{\tilde{\nu}_{I}}^{0}\phi_{\tilde{\nu}_{I}}^{0}\Big)W_{\mu}^{-}W^{+\mu} \bigg]  \nonumber \\
 & &-\frac{g^{2} + g^{\prime^{2}}}{4}\cos\theta_{W}\bigg[\Big(-1 + 
 \cos 2\theta_{W}\Big)Z_{\mu}W^{+\mu}\Big(\chi_{1}^{0}H_{2}^{1} -
\chi_{2}^{0}H_{1}^{2*} \nonumber \\
 & & + \sum_{I}\chi_{\tilde{\nu}_{I}}^{0}\tilde{L}_{2}^{I}\Big) - 
\cos\theta_{W}\sin 2\theta_{W}A_{\mu}W^{+\mu}\Big(\chi_{1}^{0}H_{2}^{1} -
\chi_{2}^{0}H_{1}^{2*} \nonumber \\
 & & + \sum_{I}\chi_{\tilde{\nu}_{I}}^{0}\tilde{L}_{2}^{I}\Big) + h.c.\bigg] + 
\frac{i(g^{2} + g^{\prime^{2}})}{4}\cos\theta_{W}\bigg[\Big(-1 + 
\cos 2\theta_{W}\Big)Z_{\mu}W^{+\mu}\Big(\phi_{1}^{0}H_{2}^{1} 
\nonumber \\
&& -\phi_{2}^{0}H_{1}^{2*} + \sum_{I}\phi_{\tilde{\nu}_{I}}^{0}\tilde{L}_{2}^{I}\Big) - 
\cos\theta_{W}\sin 2\theta_{W}A_{\mu}W^{+\mu}\Big(\phi_{1}^{0}H_{2}^{1} \nonumber \\
 & &-\phi_{2}^{0}H_{1}^{2*} + \sum_{I}\phi_{\tilde{\nu}_{I}}^{0}\tilde{L}_{2}^{I}\Big) + h.c.\bigg] - 
\frac{1}{4}(g^{2} + g^{\prime^{2}})\bigg[\sin^{2}2\theta_{W}A_{\mu}A^{\mu}\nonumber \\
 & & \Big(H_{2}^{1*}H_{2}^{1} + H_{1}^{2*}H_{1}^{2} + \sum_{I}\tilde{L}_{2}^{I*}
\tilde{L}_{2}^{I}\Big)   \nonumber \\
 & &+\cos^{2}2\theta_{W}Z_{\mu}Z^{\mu}\Big(H_{2}^{1*}H_{2}^{1} + H_{1}^{2*}H_{1}^{2} + 
\sum_{I}\tilde{L}_{2}^{I*}\tilde{L}_{2}^{I}\Big)  \nonumber \\
 & &-\sin 4\theta_{W}Z_{\mu}A^{\mu}\Big(H_{2}^{1*}H_{2}^{1} + H_{1}^{2*}H_{1}^{2} + 
\sum_{I}\tilde{L}_{2}^{I*}\tilde{L}_{2}^{I}\Big)  \nonumber \\
 & & +2\cos^{2}\theta_{W}\Big(H_{2}^{1*}H_{2}^{1} + H_{1}^{2*}H_{1}^{2} + 
\sum_{I}\tilde{L}_{2}^{I*}\tilde{L}_{2}^{I}\Big) \bigg]  \nonumber \\
 & &-\sum_{I}g^{\prime^{2}}\tilde{R}^{I*}\tilde{R}^{I}B_{\mu}B^{\mu}  \nonumber \\
 &=& -\frac{1}{4}(g^{2} + g^{\prime^{2}})\Big(\frac{1}{2}H_{i}^{0}H_{i}^{0}Z_{\mu}Z^{\mu} + 
\cos^{2}\theta_{W}H_{i}^{0}H_{i}^{0}W_{\mu}^{-}W^{+\mu}\Big)  \nonumber \\
 & &-\frac{1}{4}(g^{2} + g^{\prime^{2}})\Big(\frac{1}{2}H_{5+i}^{0}H_{5+i}^{0}Z_{\mu}Z^{\mu} 
+ \cos^{2}\theta_{W}H_{5+i}^{0}H_{5+i}^{0}
W_{\mu}^{-}W^{+\mu}\Big)  \nonumber \\
 & &+\frac{1}{4}(g^{2} + g^{\prime^{2}})\sin 2\theta_{W}\Bigg\{C_{ec}^{ij}
\bigg[\sin\theta_{W}H_{i}^{0}Z_{\mu}W^{+\mu}H_{j}^{-}   \nonumber \\
 & & + \cos\theta_{W}H_{i}^{0}A_{\mu}W^{+\mu}H_{j}^{-}\bigg] 
+ h.c. \Bigg\}  \nonumber \\
& &-\frac{i}{4}(g^{2} + g^{\prime^{2}})\sin 2\theta_{W}\Bigg\{C_{co}^{ij} 
\bigg[\sin\theta_{W}H_{5+i}^{0}Z_{\mu}W^{+\mu}H_{j}^{-}\nonumber \\
 & &  + \cos\theta_{W}H_{5+i}^{0}A_{\mu}W^{+\mu}H_{i}^{-}\bigg] - h.c. \Bigg\}  \nonumber \\
 & &-\frac{1}{4}(g^{2} + g^{\prime^{2}})\Bigg\{2\cos^{2}\theta_{W}\Big(\delta_{ij}- 
C_{c}^{ij}\Big)H_{i}^{-}
H_{j}^{+}W_{\mu}^{-}W^{+\mu} \nonumber \\
 & &+\bigg[\cos^{2}2\theta_{W}\delta_{ij} - C_{c}^{ij}\Big(4\sin^{3}\theta_{W} - 
 \cos^{2}2\theta_{W}\Big)\bigg]H_{i}^{-}H_{j}^{+}Z_{\mu}Z^{\mu}  \nonumber \\
 & & +\sin^{2}2\theta_{W}\delta_{ij}H_{i}^{-}H_{j}^{+}A_{\mu}A^{\mu} + \bigg[\sin 4\theta_{W}
\delta_{ij} \nonumber \\
 & &-C_{c}^{ij}\Big(\sin 4\theta_{W} + 
8\sin^{2}\theta_{W}\cos \theta_{W}\Big)\bigg]Z_{\mu}A^{\mu}H_{i}^{-}H_{j}^{+} \Bigg\} \hspace{1mm},
\label{vertex-ssvv}
\end{eqnarray}
where the $C_{eo}^{ij}$, $C_{co}^{ij}$ and $C_{c}^{ij}$ are defined in Eq.\ (\ref{cdef}).
The relevant Feynman rules may be summarized in Fig.\ \ref{fig1}, Fig.\ \ref{fig2}, Fig.\ \ref{fig3} 
and Fig.\ \ref{fig4}. We would emphasize some features about them. 
First, the presence of the vertices $Z_{\mu}H_{i}^{0}H_{5+j}^{0}$ 
$(i$, $j=1$, $2$, $3$, $4$, $5)$ and the forbiddance of the vertices 
$Z_{\mu}H_{i}^{0}H_{j}^{0}$ and $Z_{\mu}H_{5+i}^{0}H_{5+j}^{0}$ ($i$, 
$j=1$, $2$, $3$, $4$, $5$) couplings are determined by CP nature. 
Second, besides the $W_{\mu}^{+}Z^{\mu}H_{1}^{-}$ ($H_{1}^{-}$ is just the charged Goldstone 
boson) interaction, there are not vertices $W_{\mu}^{+}Z^{\mu}H_{i}^{-}$ 
$(i=2$, $3$, $4$, $5$, $6$, $7$, $8)$ at the tree level, 
that is the same as the MSSM with R-parity and general two-Higgs doublet models\cite{s17}.

\subsection{Self-couplings of the Higgs bosons (sleptons)}

It is straightforward to insert Eqs.\ (\ref{masseven}, \ref{ngold}, \ref{oddhiggs}, \ref{cgold}, \ref{charhiggs})
into Eqs.\ (\ref{eq-6}) to obtain the desired interaction 
terms. Similar to the interaction of gauge-Higgs (slepton) bosons, we split the Lagrangian into pieces:
\begin{equation}
{\cal L}_{int}^{S} = {\cal L}_{SSS} + {\cal L}_{SSSS}
\end{equation}
where ${\cal L}_{SSS}$ represents trilinear coupling terms, and 
${\cal L}_{SSSS}$ represents four scalar boson coupling terms. The trilinear piece
is most interesting. If the masses of the scalars are appropriate, 
the decays of one Higgs boson into two other Higgs bosons may be opened. 
After tedious computation, we have:
\begin{eqnarray}
{\cal L}_{SSS} &=& -\frac{g^{2} + g^{\prime^{2}}}{8}A_{even}^{ij}B_{even}^{k}H_{i}^{0}H_{j}^{0}H_{k}^{0} -
\frac{g^{2} + g^{\prime^{2}}}{8}A_{odd}^{ij}B_{even}^{k}H_{5+i}^{0}H_{5+j}^{0}H_{k}^{0}  \nonumber \\
 & & -A_{ec}^{kij}H_{k}^{0}
H_{i}^{-}H_{j}^{+} + iA_{oc}^{kij}H_{5+k}^{0}H_{i}^{-}H_{j}^{+}
\label{vertex-sss}
\end{eqnarray}
and
\begin{eqnarray}
{\cal L}_{SSSS} &=& -\frac{g^{2} + g^{\prime^{2}}}{32}A_{even}^{ij}A_{even}^{kl}H_{i}^{0}H_{j}^{0}H_{k}^{0}H_{l}^{0}
- \frac{g^{2} + g^{\prime^{2}}}{32}A_{odd}^{ij}A_{odd}^{kl}H_{5+i}^{0}H_{5+j}^{0}H_{5+k}^{0}H_{5+l}^{0}  \nonumber \\
& & -\frac{g^{2} + g^{\prime^{2}}}{16}A_{even}^{ij}A_{odd}^{kl}H_{i}^{0}H_{j}^{0}H_{5+k}^{0}H_{5+l}^{0} - 
{\cal A}_{ec}^{klij}H_{k}^{0}H_{l}^{0}H_{i}^{-}H_{j}^{+}  \nonumber \\
 & &-{\cal A}_{oc}^{klij}H_{5+k}^{0}H_{5+l}^{0}H_{i}^{-}H_{j}^{+} - 
i {\cal A}_{eoc}^{klij}H_{k}^{0}H_{5+l}^{0}H_{i}^{-}H_{j}^{+} - 
{\cal A}_{cc}^{klij}H_{k}^{-}H_{l}^{+}H_{i}^{-}H_{j}^{+}
\label{vertex-ssss}
\end{eqnarray}
with 
\begin{eqnarray}
A_{even}^{ij} &=& \sum_{\alpha=1}^{5}Z_{even}^{i,\alpha}Z_{even}^{j,\alpha} - 
2Z_{even}^{i,2}Z_{even}^{j,2}\hspace{1mm}, \nonumber \\
A_{odd}^{ij}  &=& \sum_{\alpha=1}^{5}Z_{odd}^{i,\alpha}Z_{odd}^{j,\alpha} 
- 2Z_{odd}^{i,2}Z_{odd}^{j,2}\hspace{1mm}, \nonumber \\
B_{even}^{i}  &=& \upsilon_{1}Z_{even}^{i,1} - \upsilon_{2}Z_{even}^{i,2} + 
\sum_{I}\upsilon_{\tilde{\nu}_{I}}Z_{even}^{i,I+2} \hspace{1mm}.
\end{eqnarray}
The definitions of $A_{ec}^{kij}$, $A_{oc}^{kij}$, ${\cal A}_{ec}^{klij}$, ${\cal A}_{oc}^{klij}$,
 ${\cal A}_{eoc}^{klij}$ and ${\cal A}_{cc}^{ijkl}$ 
can be found in Appendix.C. The Feynman rules are summarized in Fig.\ \ref{fig5} and 
Fig.\ \ref{fig6}. Note that the lepton number violation has 
led to very complicated form for the ${\cal L}_{SSS}$ and ${\cal L}_{SSSS}$.

\subsection{The couplings of Higgs to charginos (charged leptons) and neutralinos (neutrinos)}

In this subsection, we compute the interactions of the Higgs bosons with 
the supersymmetric partners of the gauge and Higgs bosons
(the gauginos and higgsinos). After spontaneous breaking of the 
gauge symmetry SU(2)$\times$U(1), the gauginos, higgsinos and 
leptons with the
same electric charge will mix as we have described in Section II. 
Let us proceed now to compute interesting interactions 
$S\tilde{\kappa}_{i}^{0}\tilde{\kappa}_{j}^{0}$ 
(Higgs-neutralinos-neutralinos interactions) etc.

The original interactions (in two-component notations) are\cite{s18}:
\begin{eqnarray}
{\cal L}_{S\kappa\kappa} &=& i\sqrt{2}g\Big(H^{1\dag}\frac{\tau^{i}}{2}\lambda_{A}^{i}\psi_{H^{1}} 
- \bar{\psi}_{H^{1}}\frac{\tau^{i}}{2}
\bar{\lambda}_{A}^{i}H^{1}\Big) - i\sqrt{2}g^{\prime}\Big(\frac{1}{2}H^{1\dag}\psi_{H^{1}}\lambda_{B} \nonumber \\
 & &- \frac{1}{2}\bar{\lambda}_{B}\bar{\psi}_{H^{1}}
H^{1}\Big) + i\sqrt{2}g\Big(H^{2\dag}\frac{\tau^{i}}{2}\lambda_{A}^{i}\psi_{H^{2}} - \bar{\psi}_{H^{2}}\frac{\tau^{i}}{2}
\bar{\lambda}_{A}^{i}H^{2}\Big) \nonumber \\
 & &+ i\sqrt{2}g^{\prime}\Big(\frac{1}{2}H^{2\dag}\psi_{H^{2}}\lambda_{B} - \frac{1}{2}\bar{\lambda}_{B}\bar{\psi}_{H^{2}}
H^{2}\Big) + i\sqrt{2}\tilde{L}^{I\dag}\Big(g\frac{\tau^{i}}{2}\lambda_{A}^{i}\psi_{L^{I}} \nonumber \\
 & &  - \frac{1}{2}g^{\prime}\lambda_{B}\psi_{L^{I}}\Big) - 
i\sqrt{2}\tilde{L}^{I}\Big(g\frac{\tau^{i}}{2}\bar{\lambda}_{A}^{i}\bar{\psi}_{L^{I}} - 
\frac{1}{2}g^{\prime}\bar{\lambda}_{B}
\bar{\psi}_{L^{I}}\Big)  \nonumber \\
 & & +i\sqrt{2}g^{\prime}\tilde{R}^{I\dag}\lambda_{B}\psi_{R^{I}} - 
i\sqrt{2}g^{\prime}\tilde{R}^{I}\bar{\lambda}_{B}\bar{\psi}_{R^{I}} - 
\frac{1}{2}l_{I}\varepsilon_{ij}\Big(\psi_{H^{1}}^{i}\psi_{L^{I}}^{j}\tilde{R}^{I}\nonumber \\
 & & + \psi_{H^{1}}^{i}\psi_{R^{I}}\tilde{R}_{j}^{I} + \psi_{R^{I}}\psi_{L^{I}}^{j}H_{i}^{1} + h.c.\Big)
\label{snn-tcom}
\end{eqnarray}
Now we sketch the derivation for the vertices, such as 
$S\tilde{\kappa}_{i}^{0}\tilde{\kappa}_{j}^{0}$ etc. Starting with the 
Eq.\ (\ref{snn-tcom}), we convert the pieces from two-component notations 
into four-component notations first, 
then using the spinor fields defined by Eq.\ (\ref{define-eneutrino}), 
Eq.\ (\ref{define-muneutrino}), Eq.\ (\ref{define-tauneutrino}), Eq.\ (\ref{define-neutralino}) 
and Eq.\ (\ref{define-chargino}), we find:
\begin{eqnarray}
{\cal L}_{S\kappa\kappa} &=& \frac{\sqrt{g^{2} + g^{\prime^{2}}}}{2}\bigg[
C_{snn}^{ij}H_{i}^{0}\bar{\kappa}_{j}^{0}P_{L}\kappa_{m}^{0}  
+C_{snn}^{ij*}H_{i}^{0}\bar{\kappa}_{j}^{0}P_{R}\kappa_{m}^{0}\bigg]  \nonumber \\
 & & + \frac{g}{\sqrt{2}}\bigg[C_{skk}^{ij}H_{i}^{0}\bar{\kappa}_{m}^{+}P_{L}\kappa_{j}^{+}
+C_{skk}^{ij*}H_{i}^{0}\bar{\kappa}_{j}^{+}P_{R}\kappa_{m}^{+}
 \bigg]  \nonumber \\
 & & + i\frac{\sqrt{g^{2} + g^{\prime^{2}}}}{2}\bigg[
C_{onn}^{ij}H_{5+i}^{0}\bar{\kappa}_{j}^{0}P_{R}\kappa_{m}^{0} 
-C_{onn}^{ij*}H_{5+i}^{0}\bar{\kappa}_{m}^{0}P_{L}\kappa_{j}^{0}\bigg] \nonumber \\
 & & +i\frac{g}{\sqrt{2}}\bigg[C_{okk}^{ij}H_{5+i}^{0}\bar{\kappa}_{m}^{+}P_{L}\kappa_{j}^{+}
 -C_{okk}^{ij*}H_{5+i}^{0}\bar{\kappa}_{m}^{+}P_{R}\kappa_{j}^{+} \bigg] \nonumber \\
&&+ \sqrt{g^{2} + g^{\prime^{2}}}\bigg[C_{Lnk}^{ij} 
\bar{\kappa}_{j}^{+}P_{L}\kappa_{m}^{0}H_{i}^{+}
-C_{Rnk}^{ij}\bar{\kappa}_{j}^{+}P_{R}\kappa_{m}^{0}H_{i}^{+}\bigg] 
\label{vertex-snn}
\end{eqnarray}
with the definitions of $C_{snn}^{ij}$, $C_{Lnk}^{ij}$, $C_{Rnk}^{ij}$ and $C_{skk}^{ij}$ are given as
\begin{eqnarray}
&&C_{snn}^{ij}=\Big(\cos\theta_{W}Z_{N}^{j,2} - \sin\theta_{W}Z_{N}^{j,1}\Big)
\Big(\sum_{\alpha=1}^{5}Z_{even}^{i,\alpha}Z_{N}^{m,2+\alpha} 
- 2Z_{even}^{i,2}Z_{N}^{m,4}\Big), \nonumber \\
&&C_{skk}^{ij}=\Big(Z_{even}^{i,1}Z_{+}^{j,1}Z_{-}^{m,2} + Z_{even}^{i,2}Z_{+}^{j,2}
Z_{-}^{m,1} + \sum_{\alpha=3}^{5}Z_{even}^{i,\alpha}Z_{+}^{j,1}Z_{-}^{m,\alpha}\Big) \nonumber \\
&&\hspace{10mm} +\frac{1}{2g}\sum_{I=1}^{3}l_{I}\Big(
Z_{even}^{i,I+2}Z_{+}^{j,I+2}Z_{-}^{m,2} - Z_{even}^{i,1}Z_{+}^{j,I+2}Z_{-}^{m,I+2}\Big), \nonumber \\
&&C_{onn}^{ij}=\Big(\cos\theta_{W}Z_{N}^{j,2} - \sin\theta_{W}Z_{N}^{j,1}\Big)\Big(
\sum_{\alpha=1}^{5}Z_{odd}^{i,\alpha}Z_{N}^{m,2+\alpha} 
- 2Z_{odd}^{i,2}Z_{N}^{m,4}\Big), \nonumber \\
&&C_{okk}^{ij}=\Big(Z_{odd}^{i,1}Z_{+}^{j,1}Z_{-}^{m,2} + Z_{odd}^{i,2}Z_{+}^{j,2}
Z_{-}^{m,1} + \sum_{\alpha=1}^{3}Z_{odd}^{i,2+\alpha}Z_{+}^{j,1}Z_{-}^{m,2+\alpha}\Big)\nonumber \\
&&\hspace{10mm} +\frac{i}{2g}\sum\limits_{I}l_{I}\Big(Z_{odd}^{i,2+I}Z_{+}^{j,2+I}Z_{-}^{m,2} 
- Z_{odd}^{i,1}Z_{+}^{j,2+I}Z_{-}^{m,2+I}\Big),  \nonumber \\
&&C_{Lnk}^{ij}=\bigg[Z_{c}^{i,1}\bigg(\frac{1}{\sqrt{2}}
\Big(\cos\theta_{W}Z_{-}^{j,2}Z_{N}^{m,2} + \sin\theta_{W}Z_{-}^{j,2}
Z_{N}^{m,1}\Big) - \cos\theta_{W}Z_{-}^{j,1}Z_{N}^{m,3}\bigg)  \nonumber \\
 & &\hspace{10mm} + \sum_{\alpha=3}^{5}Z_{c}^{i,\alpha}\bigg(\frac{1}{\sqrt{2}}
 \Big(\cos\theta_{W}Z_{-}^{j,\alpha}Z_{N}^{m,2} + 
\sin\theta_{W}Z_{-}^{j,\alpha}Z_{N}^{m,1}\Big) - 
\cos\theta_{W}Z_{-}^{j,1}Z_{N}^{m,2+\alpha}\bigg) \bigg]  \nonumber \\
 & &\hspace{10mm} + \frac{1}{2\sqrt{g^{2}+g^{\prime^{2}}}}\sum_{I=1}^{3}l_{I}\Big(Z_{c}^{i,5+I}Z_{-}^{j,2+I}Z_{N}^{m,3} 
- Z_{c}^{i,5+I}Z_{-}^{j,2}Z_{N}^{m,4+I}\Big), \nonumber \\
&&C_{Rnk}^{ij}=\bigg[Z_{c}^{i,2}\bigg(\frac{1}{\sqrt{2}}
\Big(\cos\theta_{W}Z_{+}^{*j,2}Z_{N}^{*m,2} + 
\sin\theta_{W}Z_{+}^{*j,2}Z_{N}^{*m,1}\Big)  \nonumber \\
 & &\hspace{10mm}+\cos\theta_{W}Z_{+}^{*j,1}Z_{N}^{*m,4}\bigg) 
+ \sqrt{2}\sin\theta_{W}\sum_{I=1}^{3}Z_{c}^{i,5+I}Z_{+}^{*j,2+I}
Z_{N}^{*m,1}\bigg]  \nonumber \\
 & &\hspace{10mm}+\frac{1}{2\sqrt{g^{2}+g^{\prime^{2}}}}
\sum_{I=1}^{3}l_{I}\Big(Z_{c}^{i,2+I}Z_{N}^{*m,3}Z_{+}^{*j,2+I} - Z_{c}^{i,1}
Z_{N}^{*m,3}Z_{+}^{*j,2+I}\Big). 
\label{def-coup}
\end{eqnarray}
 
Here  the project operators $P_{L,R} = \frac{1 \pm \gamma_{5}}{2}$ and the transformation 
matrices $Z_{\pm}$, $Z_{N}$ defined in Sect.II.
The corresponding Feynman rules are summarized in Fig.\ \ref{fig7}. 
As for $\kappa_{i}^{0}$ being a Majorana fermion, we note the useful
identity:
\begin{equation}
\bar{\kappa}_{j}^{0}(1 \pm \gamma_{5})\kappa_{k}^{0} = 
\bar{\kappa}_{k}^{0}(1 \pm \gamma_{5})\kappa_{j}^{0} ,
\label{majorana-protet}
\end{equation}
which holds for anticommuting four-component Majorana spinors. 
This implies that the $H_{i}^{0}\bar{\kappa}_{j}^{0}\kappa_{k}^{0}$ 
interactions can be rearranged in symmetry
under the interchange of $j$ and $k$.

Since $\nu_{e}$ $(e)$, $\nu_{\mu}$ $(\mu)$ and $\nu_{\tau}$ $(\tau)$ should be 
identified with the three lightest neutralinos (charginos) in the 
model, there must be some interesting phenomena relevant to them, such as 
$\kappa_{i}^{0}$ $(i=1,2,3,4) \rightarrow \tau H_{j}^{+}$ $(j=2$,$3$, $\cdots$, $8)$,
$\kappa_{i}^{0}$ $(i=1,2,3,4) \rightarrow \nu_{e,\mu,\tau} H_{j}^{0}$ 
$(j=1$, $2$, $\cdots$, $5)$ etc if the masses are suitable.
Namely, these interactions without R-parity conservation may 
induce interesting rare processes\cite{s20}.

\subsection{The couplings of Gauge bosons to charginos (charged leptons) and neutralinos (neutrinos)}

In this subsection we will focus on the couplings of the gauge bosons ($W$, $Z$, $\gamma$) to the charginos
(charged leptons) and neutralinos (neutrinos). 
Since we identify the three type charged leptons (three type neutrinos) 
with the three lightest charginos (neutralinos), the restrictions 
relating to them from the present experiments must
be considered carefully. The relevant interactions come from the following pieces:
\begin{eqnarray}
{\cal L}_{int}^{gcn} &=& -i\bar{\lambda}_{A}^{i}
\bar{\sigma}^{\mu}{\cal D}_{\mu}\lambda_{A}^{i} - i\bar{\lambda}_{B}\bar{\sigma}^{\mu}
{\cal D}_{\mu}\lambda_{B} - i\bar{\psi}_{H^{1}}\bar{\sigma}^{\mu}{\cal D}_{\mu}\psi_{H^{1}} \nonumber \\
 & &  - i\bar{\psi}_{H^{2}}\bar{\sigma}^{\mu}{\cal D}_{\mu}\psi_{H^{2}} 
- i\bar{\psi}_{L^{I}}\bar{\sigma}^{\mu}{\cal D}_{\mu}\psi_{L^{I}} 
- i\bar{\psi}_{R^{I}}\bar{\sigma}^{\mu}{\cal D}_{\mu}\psi_{R^{I}}
\label{lang-gcn}
\end{eqnarray}
with
\begin{eqnarray}
{\cal D}_{\mu}\lambda_{A}^{1} &=& \partial_{\mu}\lambda_{A}^{1} 
- gA_{\mu}^{2}\lambda_{A}^{3} + gA_{\mu}^{3}\lambda_{A}^{2} ,\nonumber \\
{\cal D}_{\mu}\lambda_{A}^{2} &=& \partial_{\mu}\lambda_{A}^{2} 
- gA_{\mu}^{3}\lambda_{A}^{1} + gA_{\mu}^{1}\lambda_{A}^{3} ,\nonumber \\
{\cal D}_{\mu}\lambda_{A}^{3} &=& \partial_{\mu}\lambda_{A}^{3} 
- gA_{\mu}^{1}\lambda_{A}^{2} + gA_{\mu}^{2}\lambda_{A}^{1} ,\nonumber \\
{\cal D}_{\mu}\lambda_{B} &=& \partial_{\mu}\lambda_{B} ,\nonumber \\
{\cal D}_{\mu}\psi_{H^{1}} &=& (\partial_{\mu} + igA_{\mu}^{i}\frac{\tau^{i}}{2} 
- \frac{i}{2}g^{\prime} B_{\mu})\psi_{H^{1}} ,\nonumber \\
{\cal D}_{\mu}\psi_{H^{2}} &=& (\partial_{\mu} + igA_{\mu}^{i}\frac{\tau^{i}}{2} 
+ \frac{i}{2}g^{\prime} B_{\mu})\psi_{H^{2}} ,\nonumber \\
{\cal D}_{\mu}\psi_{L^{I}} &=& (\partial_{\mu} + igA_{\mu}^{i}\frac{\tau^{i}}{2} 
- \frac{i}{2}g^{\prime} B_{\mu})\psi_{L^{I}} ,\nonumber \\
{\cal D}_{\mu}\psi_{R^{I}} &=& (\partial_{\mu} + ig^{\prime} B_{\mu})\psi_{R^{I}} .
\label{con-partial}
\end{eqnarray}
Similar to the couplings in ${\cal L}_{S\kappa\kappa}$, we convert  all spinors in
Eq.\ (\ref{lang-gcn}) into four component ones and using 
Eq.\ (\ref{define-neutralino}), Eq.\ (\ref{define-chargino}), then we obtain:
\begin{eqnarray}
{\cal L}_{int}^{gcn} &=&\Bigg\{\sqrt{g^{2}+g^{\prime^{2}}}
\sin\theta_{W}\cos\theta_{W}A_{\mu}\bar{\kappa}_{i}^{+}\gamma\kappa_{i}^{+} - 
\sqrt{g^{2}+g^{\prime^{2}}}Z_{\mu}\bar{\kappa}_{i}^{+}
\bigg[\cos^{2}\theta_{W}\delta_{ij}\gamma^{\mu}  \nonumber \\
 & &+\frac{1}{2}\Big(Z_{-}^{*i,2}Z_{-}^{j,2} + 
\sum_{I=1}^{3}Z_{-}^{*i,2+I}Z_{-}^{j,2+I}\Big)\gamma^{\mu}P_{R} \nonumber \\
&&+\Big(\frac{1}{2}Z_{+}^{*i,2}Z_{+}^{j,2} - \sum_{I=1}^{3}Z_{+}^{*i,2+I}Z_{+}^{j,2+I}
\Big)\gamma^{\mu}P_{L}\bigg]\kappa_{j}^{+} \Bigg\}  \nonumber \\
 & &+\Bigg\{ g\bar{\kappa}_{j}^{+}\bigg[ \Big(-Z_{+}^{*i,1}Z_{N}^{j,2} + \frac{1}{\sqrt{2}}Z_{+}^{*i,2}
Z_{N}^{j,4}\Big)\gamma^{\mu}P_{L} + \bigg(Z_{N}^{*i,2}Z_{-}^{j,1}  \nonumber \\
 & &+\frac{1}{\sqrt{2}}\Big(Z_{N}^{*i,3}Z_{-}^{j,2} +  \sum_{I=1}^{3}Z_{N}^{*i,4+I}Z_{-}^{j,2+I}\Big)\bigg)
\gamma^{\mu}P_{R}\bigg]\kappa_{i}^{0}W_{\mu}^{+} + h.c. \Bigg\}  \nonumber \\
 & &+\frac{\sqrt{g^{2}+g^{\prime^{2}}}}{2}\bar{\kappa}_{i}^{0}\gamma^{\mu}
\bigg[\frac{1}{2}\bigg(Z_{N}^{*i,4}Z_{N}^{j,4} - \Big(Z_{N}^{*i,3}Z_{N}^{j,3} + 
\sum_{\alpha=5}^{7}Z_{N}^{*i,\alpha}Z_{N}^{j,\alpha}\Big)\bigg)P_{L} \nonumber \\
&&- \frac{1}{2}\bigg(Z_{N}^{*i,4}Z_{N}^{j,4} - \Big(Z_{N}^{*i,3}Z_{N}^{j,3} + 
\sum_{\alpha=5}^{7}Z_{N}^{*i,\alpha}Z_{N}^{j,\alpha}\Big)\bigg)P_{R}\bigg]\kappa_{j}^{0}Z_{\mu} .
\label{vertex-gcn}
\end{eqnarray}
The corresponding Feynman rules are summarized in Fig.\ \ref{fig8}.
For we identify three lightest neutralinos (charginos) with three type
neutrinos (charged leptons), we want to emphasize some features about
Eq.\ (\ref{vertex-gcn}):
\begin{itemize}
\item For the $\gamma$ boson-$\kappa$-$\kappa$ vertices, 
there is not the lepton flavor changing current interaction at the tree level, that is same as
the SM and MSSM with R-parity.
\item For the tree level $Z$ boson-$\kappa$-$\kappa$ vertices, there are
the lepton flavor changing current interactions, this point is different
from the MSSM with R-parity.
\item Similar to the $Z$ boson-$\kappa$-$\kappa$ vertices, there are
the tree level vertices such as $W\tau\nu_{e}$ which are forbidden in the MSSM
with R-parity.
\end{itemize}

\subsection{The interactions of quarks and/or squarks with charginos (charged leptons) 
and/or neutralinos (neutrinos)}

In this subsection, we will give the Feynman rules for the interactions 
of quarks and squarks with charginos (charged leptons) and
neutralinos (neutrinos) i.e. the $\tilde{Q}q\kappa_{i}^{\pm}$ vertices. 
Because of lepton number violation so having
the mixing of neutrinos (charged leptons) and original neutralinos
(charginos), the vertices may lead to interesting phenomenology, 
thus it is interesting to write them out. 
There are two pieces contributing to the above vertices. The first 
is the supersymmetric analogue of the $q\bar{q}W^{\pm}$ 
and $q\bar{q}Z$ interaction. The second is the 
supersymmetric analogue of the $q\bar{q}H$ interaction, 
which is proportional to quark mass and depends on the properties
of the Higgs bosons in the model. These two kinds of 
vertices correspond to the terms in Eq.\ (\ref{massneutra}).

To consider the $\bar{q}q\kappa_{i}^{\pm}$ interaction first, let us write down 
the interaction in two-component spinors for fermions as the follows:
\begin{eqnarray}
{\cal L}_{\tilde{Q}q\kappa^{\pm}} &=& ig\Big(C^{IJ}\tilde{Q}_{2}^{I*}\lambda_{A}^{-}\psi_{Q_{1}}^{J} 
+ C^{IJ*}\tilde{Q}_{1}^{J*}\lambda_{A}^{+}
\psi_{Q_{2}}^{I}\Big) \nonumber \\
&&- ig\Big(C^{IJ*}\tilde{Q}_{2}^{I}\bar{\lambda}_{A}^{-}\bar{\psi}_{Q_{1}}^{J} 
+ C^{IJ}\tilde{Q}_{1}^{J}\bar{\lambda}_{A}^{+}
\bar{\psi}_{Q_{2}}^{I}\Big)  \nonumber \\
 & & -\frac{d^{I}}{2}\Big(C^{IJ}\psi_{H^{1}}^{2}\psi_{Q^{1}}^{J}\tilde{D}^{I} 
+ C^{IJ}\psi_{H^{1}}^{2}\tilde{Q}_{1}^{J}\psi_{D}^{I} +
h.c.\Big) \nonumber \\
&&+ \frac{u^{I}}{2}\Big(C^{JI*}\psi_{H^{2}}^{1}\psi_{Q^{2}}^{J}\tilde{U}^{I}  
+C^{JI*}\psi_{H^{2}}^{1}\psi_{U}^{I}\tilde{Q}_{2}^{J} + h.c.\Big) ,
\label{Qqc-tcom}
\end{eqnarray}
then convert the two-component spinors into four-component spinors as discussed above:
\begin{eqnarray}
{\cal L}_{\tilde{Q}q\kappa^{\pm}} &=& C^{IJ}\bar{\kappa}_{j}^{+}
\bigg[(-gZ_{D_{I}}^{i,1}Z_{-}^{j,1} + \frac{d^{I}}{2}Z_{D_{I}}^{i,2}
Z_{-}^{j,2})P_{L} + \frac{u^{J}}{2}Z_{+}^{j,2*}Z_{D_{I}}^{i,1}P_{R}\bigg]\psi_{u^{I}}\tilde{D}_{I,i}^{+}  \nonumber \\
 & & +C^{IJ*}\bar{\kappa}_{j}^{-}\bigg[\Big(-gZ_{U_{J}}^{i,1}Z_{+}^{j,1} + 
\frac{u^{J}}{2}Z_{U_{J}}^{i,2}Z_{+}^{j,2}\Big)P_{L}  \nonumber \\
 & &-\frac{d^{I}}{2}Z_{U_{J}}^{j,1*}Z_{-}^{j,2*}P_{R}\bigg]\psi_{d^{J}}\tilde{U}_{Ji}^{-}
+ h.c.
\label{Qqc-fcom}
\end{eqnarray}
Now $\psi_{u^{I}}$, $\psi_{d^{I}}$ are four-component quark spinors of the I-th generation. 
The $\kappa_{j}^{-}=C\bar{\kappa}_{j}^{+T}$
(C is the charge-conjugation matrix) is a charged-conjugate 
state of $\kappa_{j}^{+}$, and $\kappa_{j}^{+}$ is defined in 
Eq.\ (\ref{define-chargino}). The Feynman rules are summarized in Fig.\ \ref{fig9}. 

For the $\tilde{Q}q\kappa_{i}^{0}$ interactions, we can write the pieces in two-component notations as:
\begin{eqnarray}
{\cal L}_{\tilde{Q}q\kappa_{i}^{0}} &=& i\sqrt{2}\tilde{Q}^{I*}
\Big(g\frac{\tau^{3}}{2}\lambda_{A}^{3} + \frac{1}{6}g^{\prime}\lambda_{B}\Big)
\psi_{Q}^{I} - i\sqrt{2}\tilde{Q}^{I}\Big(g\frac{\tau^{3}}{2}\bar{\lambda}_{A}^{3} + 
\frac{1}{6}g^{\prime}\bar{\lambda}_{B}\Big)\bar{\psi}_{Q}^{I} \nonumber \\
 & & -i\frac{2\sqrt{2}}{3}g^{\prime}\tilde{U}^{I*}\lambda_{B}\psi_{U}^{I} + 
i\frac{2\sqrt{2}}{3}g^{\prime}\tilde{U}^{I}\bar{\lambda}_{B}
\bar{\psi}_{U}^{I} + i\frac{\sqrt{2}}{3}g^{\prime}\tilde{D}^{I*}\lambda_{B}\psi_{D}^{I} 
\nonumber \\
 & & - i\frac{\sqrt{2}}{3}g^{\prime}\tilde{D}^{I}
\bar{\lambda}_{B}\bar{\psi}_{D}^{I}  
+ \frac{d^{I}}{2}\bigg[\psi_{H^{1}}^{1}\psi_{Q^{2}}^{I}\tilde{D}^{I} 
+ \psi_{H^{1}}^{1}\psi_{D}^{I}\tilde{Q}_{2}^{I} + h.c.\bigg]   \nonumber \\
 & & - \frac{u^{I}}{2}\bigg[\psi_{H^{2}}^{2}\psi_{Q^{1}}^{I}\tilde{U}^{I}
+\psi_{H^{2}}^{2}\psi_{U}^{I}\tilde{Q}_{1}^{I} + h.c.\bigg]
\label{Qqn-tcom}
\end{eqnarray}
After converting Eq.\ (\ref{Qqn-tcom}) into four-component notations straightforwardly and 
using the definition for 
neutralino mass eigenstates, we have:
\begin{eqnarray}
{\cal L}_{\tilde{Q}q\kappa_{i}^{0}} &=& \kappa_{j}^{0}\Bigg\{ \bigg[\frac{e}{\sqrt{2}\sin\theta_{W}
\cos\theta_{W}}Z_{U^{I}}^{i,1*}\Big(\cos\theta_{W}
Z_{N}^{i,2} + \frac{1}{3}\sin\theta_{W}Z_{N}^{j,1}\Big)  \nonumber \\
 & &-\frac{u^{I}}{2}Z_{U^{I}}^{i,1*}Z_{N}^{j,4*}\bigg]P_{L} 
+\bigg[\frac{2\sqrt{2}}{3}g^{\prime}
Z_{U^{I}}^{i,2*}Z_{N}^{j,1} - \frac{u^{I}}{2}Z_{U^{I}}^{i,1*}Z_{N}^{j,4*}\bigg]P_{R}
\Bigg\}\psi_{u^{I}}\tilde{U}_{I,i}^{-}  \nonumber \\
 & &+\bar{\kappa}_{j}^{0}\Bigg\{\bigg[\frac{e}{\sqrt{2}\sin\theta_{W}\cos\theta_{W}}Z_{D^{I}}^{i,1}
\Big(-\cos\theta_{W}Z_{N}^{i,2} + \frac{1}{3}
\sin\theta_{W}Z_{N}^{j,1}\Big)   \nonumber \\
 & &+ \frac{d^{I}}{2}Z_{D^{I}}^{i,2}Z_{N}^{j,3}\bigg]P_{L}
+\bigg[-\frac{\sqrt{2}}{3}g^{\prime} Z_{D^{I}}^{i,2}Z_{N}^{j,1} \nonumber \\
&&+ \frac{d^{I}}{2}Z_{D^{I}}^{i,1*}Z_{N}^{j,3*}\bigg]P_{R}\Bigg\}\psi_{d^{I}}\tilde{D}_{I,i}^{+} + h.c. ,
\label{Qqn-fcom}
\end{eqnarray}
Thus the Feynman rules for the concerned interactions may be 
depicted exactly as the last two diagrams in Fig.\ \ref{fig9}.
\section{Numerical results}

In this section, we will analyze the mass spectrum of neutral Higgs, 
neutralinos and charginos numerically. We have
obtained the mass matrices by set the three type sneutrinos with non-zero 
vacuum and $\epsilon_{i} \neq 0$ $(i=1$, $2$, $3)$.
However, the matrices are too big to get the typical features. 
From now on, we will assume $\epsilon_{1} = \epsilon_{2}
=0$ and $\upsilon_{\tilde{\nu}_{e}}=\upsilon_{\tilde{\nu}_{\mu}}=0$. i.e.
only $\tau$-lepton number is violated.  We have two reasons to make the assumption:
\begin{itemize}
  \item    Under the assumption, we believe the key features will not 
  be lost but the mass matrices will turn much simple .
  \item    According to experimental indications, the 
  $\tau$-neutrino may be the heaviest among the three type neutrinos.
\end{itemize}
In the numerical calculations below, the input parameters are chosen as: 
$\alpha = \frac{e^{2}}{4\pi}=\frac{1}{128}$, $m_{Z}=91.19$GeV,
$m_{W}=80.23$GeV, $m_{\tau}=1.77$GeV, for the unknown parameters 
$m_{1}$, $m_{2}$, we assume $m_{1} = m_{2} =1000$GeV 
and the upper limit 
on $\tau$-neutrino mass $
m_{\nu_{\tau}} \leq 20$MeV is also taken into account seriously. 
Now let us consider the mass matrix 
of neutralinos first. When $\epsilon_{1}=\epsilon_{2}=0$
and $\upsilon_{\tilde{\nu}_{e}}=\upsilon_{\tilde{\nu}_{\mu}}=0$, 
the Eq.\ (\ref{neutralino-matrix}) is simplified as:
\begin{equation}
{\cal M}_{N} = \left(
\begin{array}{ccccc}
2m_{1} & 0 & -\frac{1}{2}g^{\prime}\upsilon_{1} & 
\frac{1}{2}g^{\prime}\upsilon_{2} & -\frac{1}{2}g^{\prime}\upsilon_{\tilde{\nu}_{\tau}} \\
0 & 2m_{2} & \frac{1}{2}g\upsilon_{1} & -\frac{1}{2}g\upsilon_{2} & 
\frac{1}{2}g\upsilon_{\tilde{\nu}_{\tau}} \\
-\frac{1}{2}g^{\prime}\upsilon_{1} & \frac{1}{2}g\upsilon_{1} & 0 & -\frac{1}{2}\mu & 0 \\
\frac{1}{2}g^{\prime}\upsilon_{2} & -\frac{1}{2}g\upsilon_{2} & 
-\frac{1}{2}\mu & 0 & \frac{1}{2}\epsilon_{3} \\
-\frac{1}{2}g^{\prime}\upsilon_{\tilde{\nu}_{\tau}} & 
\frac{1}{2}g\upsilon_{\tilde{\nu}_{\tau}} & 0 & \frac{1}{2}\epsilon_{3} & 0 
\end{array} \right)
\label{rneutralino-matrix}
\end{equation}
As stated above, a strong restriction imposes on the matrix is from $m_{\nu_{\tau}} \leq 20$MeV.
Ref\cite{s21} has discussed this limit impacting on the parameter space,
the numerical result indicates the large value of $\epsilon_{3}$ cannot be ruled out. In order to show
the problem precisely, let us consider the equation for the eigenvalues of Eq.\ (\ref{rneutralino-matrix}):
\begin{eqnarray}
Det(\lambda - {\cal M}_{N}) &=& \lambda^{5} - 2(m_{1} + m_{2}) \lambda^{4} + \Big(4m_{1}m_{2} -\frac{1}{4}
(\epsilon_{3}^{2} + \mu^{2}) - M_{Z}^{2}\Big) \lambda^{3}  \nonumber \\
 & &+\bigg[\frac{1}{2}(m_{1} + m_{2})(\epsilon_{3}^{2} +
\mu^{2}) +   2(m_{1} + m_{2})M_{Z}^{2} \nonumber \\
&& + \frac{1}{4}(g^{2} + g^{\prime^{2}})\upsilon_{2}(-\mu\upsilon_{1} + 
\epsilon_{3}\upsilon_{\tilde{\nu}_{\tau}}) \bigg] \lambda^{2}  \nonumber \\
 & &+\bigg[-m_{1}m_{2}(\mu^{2} + \epsilon_{3}^{2}) + \frac{1}{16}(g^{2} + 
g^{\prime^{2}})(\epsilon_{3}\upsilon_{1} + \mu\upsilon_{\tilde{\nu}_{\tau}})^{2} \nonumber \\
 & & + \frac{1}{2}(g^{2}m_{1} +g^{\prime^{2}}m_{2})(\mu\upsilon_{1}\upsilon_{2} - 
\epsilon_{3}\upsilon_{2}\upsilon_{\tilde{\nu}_{\tau}})\bigg] \lambda \nonumber \\
&&- \frac{1}{8}(g^{2}m_{1} + g^{\prime^{2}}m_{2})\Big(\mu\upsilon_{\tilde{\nu}_{\tau}} +
\epsilon_{3}\upsilon_{1}\Big)^{2}  \nonumber \\
 &=& \lambda^{5} + {\cal A}_{N}\lambda^{4} + {\cal B}_{N}\lambda^{3} + {\cal C}_{N}\lambda^{2} + {\cal D}_{N}\lambda  
 + {\cal E}_{N}. 
\label{eigen-neutralino}
\end{eqnarray}
For further discussions, let us introduce new symbols $X$, $Y$ as:
\begin{eqnarray}
X &=& \epsilon_{3}\cos\theta_{\upsilon} + \mu\sin\theta_{\upsilon}, \nonumber \\
Y &=& -\epsilon_{3}\sin\theta_{\upsilon} + \mu\cos\theta_{\upsilon} ,
\label{define-XY}
\end{eqnarray}
thus we have the coefficients of Eq.\ (\ref{eigen-neutralino}):
\begin{eqnarray}
{\cal A}_{N} &=& - 2(m_{1} + m_{2}), \nonumber \\
{\cal B}_{N} &=& -\frac{1}{4}(X^{2} + Y^{2}) + 4m_{1}m_{2} - M_{Z}^{2}, \nonumber \\
{\cal C}_{N} &=& \frac{1}{2}(m_{1} + m_{2})(X^{2} + Y^{2}) + 2(m_{1} + m_{2})M_{Z}^{2} -  
M_{Z}^{2}\cos\beta\sin\beta Y,   \nonumber \\
{\cal D}_{N} &=& -m_{1}m_{2}(X^{2} + Y^{2}) + \frac{1}{4}M_{Z}^{2}\cos^{2}\beta
X^{2} + \frac{1}{2}(g^{2}m_{1} + g^{\prime^{2}})\upsilon^{2}\sin\beta\cos\beta Y , \nonumber \\
{\cal E}_{N} &=& -\frac{1}{8}(g^{2}m_{1} + g^{\prime^{2}}m_{2})\upsilon^{2}\cos^{2}\beta X^{2}
\label{coeffi1}
\end{eqnarray}
If we fix the $\tau$-neutrino mass $m_{\nu_{\tau}}$ as an input parameter, so the 
equation $Det(\lambda - {\cal M}_{N})=0$ can be written as:
\begin{equation}
\Big(\lambda - m_{\nu_{e}}\Big)\Big(\lambda^{4} + {\cal A}^{\prime}_{N}\lambda^{3} 
+ {\cal B}^{\prime}_{N}\lambda^{2} + {\cal C}^{\prime}_{N}\lambda + 
{\cal D}^{\prime}_{N}\Big) = 0.
\label{eigen-neutralino1}
\end{equation}
The coefficients ${\cal A}^{\prime}_{N}$, ${\cal B}^{\prime}_{N}$, 
${\cal C}^{\prime}_{N}$, ${\cal D}^{\prime}_{N}$ are related to the 
"original" ones ${\cal A}_{N}$, ${\cal B}_{N}$, ${\cal C}_{N}$, 
${\cal D}_{N}$, and ${\cal E}_{N}$ as: 
\begin{eqnarray}
{\cal A}^{\prime}_{N} &=& {\cal A}_{N} + m_{\nu_{\tau}}, \nonumber \\
{\cal B}^{\prime}_{N} &=& {\cal B}_{N} + m_{\nu_{\tau}}{\cal A}^{\prime}_{N}, \nonumber \\
{\cal C}^{\prime}_{N} &=& {\cal C}_{N} + m_{\nu_{\tau}}{\cal B}^{\prime}_{N}, \nonumber \\
{\cal D}^{\prime}_{N} &=& {\cal D}_{N} + m_{\nu_{\tau}}{\cal C}^{\prime}_{N} = -\frac{{\cal E}_{N}}{m_{\nu_{\tau}}}.
\label{coeffi2}
\end{eqnarray}
To obtain the masses of the other four neutralinos, let us solve the Eq.\ (\ref{eigen-neutralino1}) 
by the numerical method.
In Fig.\ \ref{fig10}, we plot the mass of the lightest neutralino versus X. The three lines correspond to $m_{
\nu_{\tau}} = 20$MeV, $2$MeV and $0.2$MeV respectively. From the figure, we find that the curve corresponding to
$m_{\nu_{\tau}}=20$MeV is the lowest and the second low one is correspond to
$m_{\nu_{\tau}}=2$MeV, so the tendency is that the curves 
are going "up" as the $\tau$-neutrino mass is decreasing. 
If the mass of the lightest neutralino is not too heavy (such as
$m_{\kappa_{0}^{1}} \leq 300$GeV), the absolute value of X cannot take very large (for example
$|X| \leq 800$GeV).

As for the mass of the charginos, when $\epsilon_{1} = \epsilon_{2} =0$, and $\upsilon_{\tilde{\nu}_{e}} = 
\upsilon_{\tilde{\nu}_{\mu}}=0$, the Eq.\ (\ref{chargino-matrix}) becomes:
\begin{equation}
{\cal M}_{C} = \left(
\begin{array}{ccc}
2m_{2} & \frac{e\upsilon_{2}}{\sqrt{2}S_{W}} & 0  \\
\frac{e\upsilon_{1}}{\sqrt{2}S_{W}} & \mu  & \frac{l_{3}\upsilon_{\tilde{\nu}_{\tau}}}{\sqrt{2}} \\
\frac{e\upsilon_{\tilde{\nu}_{\tau}}}{\sqrt{2}S_{W}} & \epsilon_{3} & \frac{l_{3}\upsilon_{1}}{\sqrt{2}}
\end{array}  \right)
\label{rchargino-matrix}
\end{equation}
Because $m_{\tau}^{2}$ should be the lightest eigenvalue of the 
matrix ${\cal M}_{C}^{\dag}{\cal M}_{C}$, after taking the eigenvalue $m_{\tau}^{2}$ away, 
the surviving eigenvalue equation becomes:
\begin{equation}
\lambda^{2} - {\cal A}_{c}\lambda + {\cal B}_{C} = 0.
\end{equation}
Here,
\begin{eqnarray}
{\cal A}_{C} &=& X^{2} + Y^{2} + 4m_{2}^{2} + l_{3}^{2}\frac{\upsilon_{1}^{2} + \upsilon_{\tilde{\nu}_{\tau}}^{2}}{2}
 + \frac{e^{2}\upsilon^{2}}{2S_{W}^{2}}  \nonumber \\
{\cal B}_{C} &=& \frac{2l_{3}^{2}}{m_{\tau}^{2}} \Bigg\{m_{2}\upsilon\cos\beta Y + \frac{e^{2}}{4S_{W}^{2}}
\upsilon^{3}\cos^{2}\beta\sin\beta \Big(\sin^{2}\theta_{\upsilon} - \cos^{2}\theta_{\upsilon}\Big) \Bigg\}^{2}
\label{coeffi3}
\end{eqnarray}
with the parameters $X$, $Y$ are defined by Eq.\ (\ref{define-XY}). Therefore the masses 
of the other two charginos are expressed as:
\begin{equation}
m_{\kappa_{1,2}^{\pm}}^{2} = \frac{1}{2}\bigg\{ {\cal A}_{C} \mp \sqrt{{\cal A}_{C}^{2} - 4{\cal B}_{C}}\bigg\}.
\label{charginomass}
\end{equation}
The parameter $l_{3}$ can be fixed by the condition $Det|m_{\tau}^{2} - {\cal M}_{c}^{\dag}{\cal M}_{c}|=0$.
In Fig.\ \ref{fig11}, we plot the mass of the lightest chargino versus X. The three lines correspond to
$m_{\nu_{\tau}} = 20$MeV, $2$MeV and $0.2$MeV respectively. Similar to the neutralinos, we find that
the curve corresponding to $m_{\nu_{\tau}} = 20$MeV is the lowest, the second low one 
is correspond to $m_{\nu_{\tau}} = 2$MeV and the tendency is very similar 
to the case for neutralinos. 
This can be understood as following: when the values of $m_{1}$, $m_{2}$, $\tan\beta$, $\tan\theta_{\upsilon}$
and X are fixed, the value of Y will be fixed by the mass of $\tau$-neutrino. In the numerical computation, we
find that the absolute value of Y turns small, as the $m_{\nu_{\tau}}$ changes large. 
This is the reason why the curve
corresponding to $m_{\nu_{\tau}} = 20$MeV is the lowest among the three curves which we have computed here.

Now, we turn to discuss the mass matrix of the neutral Higgs. Under the same assumption, the mass matrix for 
CP-even Higgs reduces to:
\begin{equation}
{\cal M}_{even}^{2} = \left(
\begin{array}{ccc}
r_{11} & -\frac{g^{2} + g^{\prime^{2}}}{4}\upsilon_{1}\upsilon_{2} - B\mu  &  \frac{g^{2} + g^{\prime^{2}}}{4}
\upsilon_{1}\upsilon_{\tilde{\nu}_{\tau}} - \mu\epsilon_{3}  \\
-\frac{g^{2} + g^{\prime^{2}}}{4}\upsilon_{1}\upsilon_{2} - B\mu & r_{22} & -\frac{g^{2} + g^{\prime^{2}}}{4}
\upsilon_{2}\upsilon_{\tilde{\nu}_{\tau}} + B_{3}\epsilon_{3}  \\
\frac{g^{2} + g^{\prime^{2}}}{4}\upsilon_{1}\upsilon_{\tilde{\nu}_{\tau}} - \mu\epsilon_{3} & -\frac{g^{2} + g^{\prime^{2}}}{4}
\upsilon_{2}\upsilon_{\tilde{\nu}_{\tau}} + B_{3}\epsilon_{3} & r_{33}
\end{array}
\right)
\label{rmatrix-even}
\end{equation}
with
\begin{eqnarray}
r_{11} &=& \frac{g^{2} + g^{\prime^{2}}}{8}(3\upsilon_{1}^{2} - \upsilon_{2}^{2} + \upsilon_{\tilde{\nu}_{\tau}}^{2}) +
|\mu|^{2} + m_{H^{1}}^{2}  \nonumber \\
 &=& \frac{g^{2} + g^{\prime^{2}}}{4}\upsilon_{1}^{2} + \mu\epsilon_{3}\frac{\upsilon_{\tilde{\nu}_{\tau}}}{\upsilon_{1}}
+ B\mu\frac{\upsilon_{2}}{\upsilon_{1}}  \nonumber \\
r_{22} &=& \frac{g^{2} + g^{\prime^{2}}}{8}(-\upsilon_{1}^{2} + 3\upsilon_{2}^{2} - \upsilon_{\tilde{\nu}_{\tau}}^{2}) +
|\mu|^{2} + |\epsilon_{3}|^{2} + m_{H^{2}}^{2} \nonumber \\
 &=& \frac{g^{2} + g^{\prime^{2}}}{4}\upsilon_{2}^{2} + B\mu\frac{\upsilon_{1}}{\upsilon_{2}}
- B_{3}\epsilon_{3}\frac{\upsilon_{\tilde{\nu}_{\tau}}}{\upsilon_{2}}  \nonumber \\
r_{33} &=& \frac{g^{2} + g^{\prime^{2}}}{8}(\upsilon_{1}^{2} - \upsilon_{2}^{2} + 3\upsilon_{\tilde{\nu}_{\tau}}^{2}) +
|\epsilon_{3}|^{2} + m_{L^{3}}^{2}  \nonumber \\
 &=& \frac{g^{2} + g^{\prime^{2}}}{4}\upsilon_{\tilde{\nu}_{\tau}}^{2} + \mu\epsilon_{3}\frac{\upsilon_{1}}{\upsilon_{
\tilde{\nu}_{\tau}}}
- B_{3}\epsilon_{3}\frac{\upsilon_{2}}{\upsilon_{\tilde{\nu}_{\tau}}} 
\label{r123}
\end{eqnarray}
The mass matrix of CP-odd Higgs reduces to: 
\begin{equation}
{\cal M}_{odd}^{2} =
\left(
\begin{array}{ccc}
s_{11} & B\mu & -\mu\epsilon_{3}  \\
B\mu & s_{22} & -B_{3}\epsilon_{3} \\
-\mu\epsilon_{3} & -B_{3}\epsilon_{3} & s_{33}
\end{array}
\right)  \label{rmassodd}
\end{equation}
with
\begin{eqnarray}
s_{11} &=& \frac{g^{2} + g^{\prime^{2}}}{8}(\upsilon_{1}^{2} - \upsilon_{2}^{2} + 
\upsilon_{\tilde{\nu}_{\tau}}^{2}) + |\mu|^{2}
+ m_{H^{1}}^{2}  ,\nonumber \\
 &=& \mu\epsilon_{3}\frac{\upsilon_{\tilde{\nu}_{\tau}}}{\upsilon_{1}} + 
 B\mu\frac{\upsilon_{2}}{\upsilon_{1}} \nonumber \\
s_{22} &=& -\frac{g^{2} + g^{\prime^{2}}}{8}(\upsilon_{1}^{2} - \upsilon_{2}^{2} 
+ \upsilon_{\tilde{\nu}_{\tau}}^{2}) + |\mu|^{2}
+|\epsilon_{3}|^{2} + m_{H^{2}}^{2}  \nonumber \\
 &=& B\mu\frac{\upsilon_{1}}{\upsilon_{2}} - 
 B_{3}\epsilon_{3}\frac{\upsilon_{\tilde{\nu}_{\tau}}}{\upsilon_{2}} ,\nonumber \\
s_{33} &=& \frac{g^{2} + g^{\prime^{2}}}{8}(\upsilon_{1}^{2} - \upsilon_{2}^{2} + \upsilon_{\tilde{\nu}_{\tau}}^{2}) + 
|\epsilon_{3}|^{2} + m_{L^{3}}^{2} \nonumber \\
 &=& \mu\epsilon_{3}\frac{\upsilon_{1}}{\upsilon_{\tilde{\nu}_{\tau}}} - B_{3}\epsilon_{3}\frac{\upsilon_{2}}{\upsilon_{
\tilde{\nu}_{\tau}}}.
\label{s123}
\end{eqnarray}
Introducing the following variables:
\begin{eqnarray}
X_{s} &=& B\mu ,\nonumber \\
Y_{s} &=& \mu\epsilon_{3} ,\nonumber \\
Z_{s} &=& B_{3}\epsilon_{3},
\label{xyzs}
\end{eqnarray}
the masses of the neutral Higgs can be determined from the $X_{s}$, $Y_{s}$, $Z_{s}$ 
and $\tan\beta$, $\tan\theta_{\upsilon}$.
For the masses of CP-odd Higgs, we define:
\begin{eqnarray}
{\cal A} &=& X_{s} \bigg(\frac{\upsilon_{1}}{\upsilon_{2}} + 
\frac{\upsilon_{2}}{\upsilon_{1}}\bigg) + Y_{s} \bigg(\frac{
\upsilon_{1}}{\upsilon_{\tilde{\nu}_{\tau}}} + \frac{\upsilon_{\tilde{\nu}_{\tau}}}{\upsilon_{1}}\bigg) 
- Z_{s}\bigg(\frac{\upsilon_{2}}{
\upsilon_{\tilde{\nu}_{\tau}}} + \frac{\upsilon_{\tilde{\nu}_{\tau}}}{\upsilon_{2}}\bigg),  \nonumber \\
{\cal B} &=& - Y_{s}Z_{s} \bigg(\frac{\upsilon_{1}}{\upsilon_{2}} + \frac{\upsilon_{2}}{\upsilon_{1}}\bigg) -
X_{s}Z_{s}\bigg(\frac{\upsilon_{1}}{\upsilon_{\tilde{\nu}_{\tau}}} + 
\frac{\upsilon_{\tilde{\nu}_{\tau}}}{\upsilon_{1}}\bigg) + X_{s}Y_{s}
\bigg(\frac{\upsilon_{2}}{\upsilon_{\tilde{\nu}_{\tau}}} + 
\frac{\upsilon_{\tilde{\nu}_{\tau}}}{\upsilon_{2}}\bigg)  \nonumber \\
 & & +X_{s}Y_{s}\frac{\upsilon_{1}^{2}}{\upsilon_{2}\upsilon_{\tilde{\nu}_{\tau}}} - X_{s}Z_{s}\frac{\upsilon_{
2}^{2}}{\upsilon_{1}\upsilon_{\tilde{\nu}_{\tau}}} - Y_{s}Z_{s} 
\frac{\upsilon_{\tilde{\nu}_{\tau}}^{2}}{\upsilon_{1}\upsilon_{2}},
\label{calab}
\end{eqnarray}
the masses of the two CP-odd Higgs can be given as:
\begin{equation}
m_{H_{3+2,3}^{0}}^{2} = \frac{1}{2}\bigg({\cal A} \mp \sqrt{{\cal A}^{2} - 4 {\cal B}} \bigg).
\label{massodd1}
\end{equation}
In Fig.\ \ref{fig12}, we plot the mass of the lightest CP-odd Higgs versus the mass of the lightest CP-even Higgs,
where the ranges of the parameters are: $ -10^{5}$GeV $^{2}$ $ \leq X_{s}, Y_{s}, Z_{s} \leq 10^{5} $GeV$^{2}$ and $.5 \leq \tan\beta,
\tan\theta_{\upsilon} \leq 50$. From the Fig.\ \ref{fig12}, we can find that 
there are no limit on the $m_{H_{5}^{0}}$ when
we change those parameters in the above ranges. As for the lightest CP-even Higgs, 
the difference from the MSSM with R-parity is that
$m_{H_{1}^{0}}$ can larger than $m_{Z}$ at the tree level. 
This can be understood from Eq.\ (\ref{bound-masshiggs}), under the assumptions, 
we have
\begin{equation}
m_{H_{1}^{0}}^{2} \leq m_{H_{3}^{0}} m_{Z}\cos 2\beta \frac{1-\frac{1}{2}\frac{m_{Z}^{2}}{m_{H_{3}^{0}}^{2}}}{1-\frac{1}{2}
\bigg(\frac{m_{Z}^{2}\cos^{2}2\beta}{m_{H_{3}^{0}}^{2}}\bigg)^{\frac{1}{2}}}
\end{equation}
where the $m_{H_{3}^{0}}$ is the mass of the heaviest CP-even Higgs in this case and we 
cannot give the stringent limit on it as in the MSSM with R-parity.

In summary, we have analyzed the mass spectrum in the 
MSSM with bilinear R-parity violation. From the restriction $m_{\nu_{\tau}} \leq 20$MeV, 
we cannot rule out the possibilities with large $\epsilon_{3}$ and 
$\upsilon_{\tilde{\nu}_{\tau}}$. We also derived the Feynman rules in the $\prime$t-Hooft
Feynman gauge, which are convenient when we study the phenomenology 
beyond the tree level in the model. Recent experimental signals of
neutrino masses and mixing may provide the first glimpses of the lepton number violation effects,
Ref\cite{sh1} have study the Neutrino Oscillations experiment constraint on the 
parameter space of the model. Considering both the fermionic and scalar sectors,
they find that a large area of the parameter space is allowed.
Here, we would also like to  point out
some references have analyzed the $0\nu \beta\beta$-decay
in the model\cite{s22} and obtained new stringent upper limits on the first 
generation R-parity violating parameters, $\epsilon_{1}$ and $\upsilon_{\tilde{\nu}_{e}}$; 
whereas for the other two generations, there are not very serious restrictions on the upper limits of
the R-parity violating parameters. As for other interesting processes in the model, they
are discussed by Ref\cite{sh2}.

{\Large\bf Acknowledgment} This work was supported in part by the National Natural
Science Foundation of China and the Grant No. LWLZ-1298 of the Chinese Academy of
Sciences.

\appendix

\section{The mass matrix of charged Higgs}

In the case of charged Higgs, with the current basis 
$\Phi_{c}=(H_{2}^{1*}$, $H_{1}^{2}$, $\tilde{L}_{2}^{1*}$,  
$\tilde{L}_{2}^{2*}$, $\tilde{L}_{2}^{3*}$, $\tilde{R}^{1}$, $\tilde{R}^{2}$, $\tilde{R}^{3})$,
the symmetric matrix ${\cal M}_{c}^{2}$ is given as follows:
\begin{eqnarray}
{\cal M}_{c 1,1}^{2} & = & \frac{g^{2}}{4}\upsilon_{1}^{2} - \frac{g^{2}-g'^{2}}{8}(
\upsilon_{1}^{2}-\upsilon_{2}^{2} + \sum_{I}\upsilon_{\tilde{\nu}_{I}}^{2}) + \mu^{2} + \sum_{I}\frac{
1}{2}l_{I}^{2}\upsilon_{\tilde{\nu}_{I}}^{2} + m_{H^{1}}^{2}  \nonumber  \\
   & = & \frac{g^{2}}{4}(\upsilon_{2}^{2}-\sum_{I}\upsilon_{\tilde{\nu}_{I}}^{2}) + \sum_{I}\frac{1}{2}l_{I}^{2}
\upsilon_{\tilde{\nu}_{I}}^{2}
 +\sum_{I}\mu\epsilon_{I}\frac{\upsilon_{\tilde{\nu}_{I}}}{\upsilon_{1}}+B\mu\frac{\upsilon_{2}}{
\upsilon_{1}}, \nonumber  \\
{\cal M}_{c 1,2}^{2} &=& \frac{g^{2}}{4}\upsilon_{1}\upsilon_{2} + B\mu , \nonumber \\
{\cal M}_{c 1,3}^{2} &=& \frac{g^{2}}{4}\upsilon_{1}\upsilon_{\tilde{\nu}_{e}} -\mu\epsilon_{1}-\frac{
1}{2}l_{1}^{2}\upsilon_{1}\upsilon_{\tilde{\nu}_{e}}  , \nonumber \\
{\cal M}_{c 1,4}^{2} &=& \frac{g^{2}}{4}\upsilon_{1}\upsilon_{\tilde{\nu}_{\mu}} -\mu\epsilon_{2}-\frac{
1}{2}l_{2}^{2}\upsilon_{1}\upsilon_{\tilde{\nu}_{\mu}}  , \nonumber \\
{\cal M}_{c 1,5}^{2} &=& \frac{g^{2}}{4}\upsilon_{1}\upsilon_{\tilde{\nu}_{\tau}} -\mu\epsilon_{3}-\frac{
1}{2}l_{3}^{2}\upsilon_{1}\upsilon_{\tilde{\nu}_{\tau}}  , \nonumber \\
{\cal M}_{c 1,6}^{2} &=& \frac{1}{\sqrt{2}}l_{1}\epsilon_{1}\upsilon_{2} +
l_{s_{1}}\frac{\mu \upsilon_{\tilde{\nu}_{e}}}{\sqrt{2}} , \nonumber \\
{\cal M}_{c 1,7}^{2} &=& \frac{1}{\sqrt{2}}l_{2}\epsilon_{2}\upsilon_{2} +
l_{s_{2}}\frac{\mu \upsilon_{\tilde{\nu}_{\mu}}}{\sqrt{2}} , \nonumber \\
{\cal M}_{c 1,8}^{2} &=& \frac{1}{\sqrt{2}}l_{3}\epsilon_{3}\upsilon_{2} +
l_{s_{3}}\frac{\mu \upsilon_{\tilde{\nu}_{\tau}}}{\sqrt{2}} , \nonumber \\
{\cal M}_{c 2,2}^{2} &=& \frac{g^{2}}{4}\upsilon_{2}^{2} + \frac{1}{8}(g^{2}-g'^{2})
(\upsilon_{1}^{2}-\upsilon_{2}^{2}+\sum_{I}\upsilon_{\tilde{\nu}_{I}}^{2}) + \mu^{2} + 
\sum_{I}\epsilon_{I}^{2} + m_{H^{2}}^{2}   \nonumber  \\
 & = & \frac{g^{2}}{4}(\upsilon_{1}^{2}+\sum_{I}\upsilon_{\tilde{\nu}_{I}}^{2}) - \sum_{I}B_{I}\epsilon_{I}
\frac{\upsilon_{\tilde{\nu}_{I}}}{\upsilon_{2}} + B\mu\frac{\upsilon_{1}}{\upsilon_{2}}, \nonumber \\
{\cal M}_{c 2,3}^{2} &=& \frac{g^{2}}{4}\upsilon_{2}\upsilon_{\tilde{\nu}_{e}} - B_{1}\epsilon_{1} ,\nonumber  \\
{\cal M}_{c 2,4}^{2} &=& \frac{g^{2}}{4}\upsilon_{2}\upsilon_{\tilde{\nu}_{\mu}} - B_{2}\epsilon_{2} ,\nonumber  \\
{\cal M}_{c 2,5}^{2} &=& \frac{g^{2}}{4}\upsilon_{2}\upsilon_{\tilde{\nu}_{\tau}} - B_{3}\epsilon_{3} ,\nonumber  \\
{\cal M}_{c 2,6}^{2} &=& \frac{l_{1}}{\sqrt{2}}\mu\upsilon_{\tilde{\nu}_{e}} + \frac{
l_{1}}{\sqrt{2}}\epsilon_{1}\upsilon_{1} ,\nonumber \\
{\cal M}_{c 2,7}^{2} &=& \frac{l_{2}}{\sqrt{2}}\mu\upsilon_{\tilde{\nu}_{\mu}} + \frac{
l_{2}}{\sqrt{2}}\epsilon_{2}\upsilon_{1} ,\nonumber \\
{\cal M}_{c 2,8}^{2} &=& \frac{l_{3}}{\sqrt{2}}\mu\upsilon_{\tilde{\nu}_{\tau}} + \frac{
l_{3}}{\sqrt{2}}\epsilon_{3}\upsilon_{1} ,\nonumber \\
{\cal M}_{c 3,3}^{2} &=& \frac{g^{2}}{4}\upsilon_{\tilde{\nu}_{e}}^{2} - \frac{1}{8}(g^{2}-g'^{2})
(\upsilon_{1}^{2}-\upsilon_{2}^{2}+\sum_{I}\upsilon_{\tilde{\nu}_{I}}^{2}) +\epsilon_{1}^{2} + 
\frac{l_{1}^{2}}{2}\upsilon_{1}^{2} + m_{L^{1}}^{2}  \nonumber \\
 & = & \frac{g^{2}}{4}(\upsilon_{2}^{2}-\upsilon_{1}^{2})+\epsilon_{1}\frac{\mu\upsilon_{1}}{
\upsilon_{\tilde{\nu}_{e}}} - B_{1}\frac{\epsilon_{1}\upsilon_{2}}{\upsilon_{\tilde{\nu}_{e}}} + 
\frac{l_{1}^{2}}{2}\upsilon_{1}^{2}  \nonumber \\
 & & -\epsilon_{1}\epsilon_{2}\frac{\upsilon_{\tilde{\nu}_{\mu}}}{\upsilon_{\tilde{\nu}_{e}}} - 
\epsilon_{1}\epsilon_{3}\frac{\upsilon_{\tilde{\nu}_{\tau}}}{\upsilon_{\tilde{\nu}_{e}}} - 
\frac{g^{2}}{4}(\upsilon_{\tilde{\nu}_{\mu}}^{2}+\upsilon_{\tilde{\nu}_{\tau}}^{2}), \nonumber \\
{\cal M}_{c 3,4}^{2} &=& \frac{g^{2}}{4}\upsilon_{\tilde{\nu}_{e}}\upsilon_{\tilde{\nu}_{\mu}} 
+ \epsilon_{1}\epsilon_{2} ,\nonumber \\
{\cal M}_{c 3,5}^{2} &=& \frac{g^{2}}{4}\upsilon_{\tilde{\nu}_{e}}\upsilon_{\tilde{\nu}_{\tau}} 
+ \epsilon_{1}\epsilon_{3} ,\nonumber \\
{\cal M}_{c 3,6}^{2} &=& \frac{1}{\sqrt{2}}l_{1}\mu\upsilon_{2}+\frac{1}{
\sqrt{2}}l_{s_{1}}\mu\upsilon_{1} , \nonumber  \\
{\cal M}_{c 3,7}^{2} &=& 0 ,\nonumber \\
{\cal M}_{c 3,8}^{2} &=& 0 ,\nonumber \\
{\cal M}_{c 4,4}^{2} &=& \frac{g^{2}}{4}\upsilon_{\tilde{\nu}_{\mu}}^{2} - \frac{1}{8}(g^{2}-g'^{2})
(\upsilon_{1}^{2}-\upsilon_{2}^{2}+\sum_{I}\upsilon_{\tilde{\nu}_{I}}^{2}) +\epsilon_{2}^{2} + 
\frac{l_{2}^{2}}{2}\upsilon_{1}^{2} + m_{L^{2}}^{2}  \nonumber \\
 & = & \frac{g^{2}}{4}(\upsilon_{2}^{2}-\upsilon_{1}^{2})+\epsilon_{2}\frac{\mu\upsilon_{1}}{
\upsilon_{\tilde{\nu}_{\mu}}} - B_{2}\frac{\epsilon_{2}\upsilon_{2}}{\upsilon_{\tilde{\nu}_{\mu}}} + 
\frac{l_{2}^{2}}{2}\upsilon_{1}^{2}  \nonumber \\
 & & -\epsilon_{1}\epsilon_{2}\frac{\upsilon_{\tilde{\nu}_{e}}}{\upsilon_{\tilde{\nu}_{\mu}}} - 
\epsilon_{2}\epsilon_{3}\frac{\upsilon_{\tilde{\nu}_{\tau}}}{\upsilon_{\tilde{\nu}_{\mu}}} - 
\frac{g^{2}}{4}(\upsilon_{\tilde{\nu}_{e}}^{2}+\upsilon_{\tilde{\nu}_{\tau}}^{2}), \nonumber \\
{\cal M}_{c 4,5}^{2} &=& \frac{g^{2}}{4}\upsilon_{\tilde{\nu}_{\mu}}\upsilon_{\tilde{\nu}_{\tau}} ,\nonumber \\
{\cal M}_{c 4,6}^{2} &=& 0  ,\nonumber \\
{\cal M}_{c 4,7}^{2} &=& \frac{1}{\sqrt{2}}l_{2}\mu\upsilon_{2}-\frac{1}{
\sqrt{2}}l_{s_{2}}\mu\upsilon_{1} , \nonumber  \\
{\cal M}_{c 4,8}^{2} &=& 0 ,\nonumber \\
{\cal M}_{c 5,5}^{2} &=& \frac{g^{2}}{4}\upsilon_{\tilde{\nu}_{\tau}}^{2} - \frac{1}{8}(g^{2}-g'^{2})
(\upsilon_{1}^{2}-\upsilon_{2}^{2}+\sum_{I}\upsilon_{\tilde{\nu}_{I}}^{2}) +\epsilon_{3}^{2} + 
\frac{l_{3}^{2}}{2}\upsilon_{1}^{2} + m_{L^{3}}^{2}  \nonumber \\
 & = & \frac{g^{2}}{4}(\upsilon_{2}^{2}-\upsilon_{1}^{2})+\epsilon_{3}\frac{\mu\upsilon_{1}}{
\upsilon_{\tilde{\nu}_{\tau}}} - B_{3}\frac{\epsilon_{3}\upsilon_{2}}{\upsilon_{\tilde{\nu}_{\tau}}} + 
\frac{l_{3}^{2}}{2}\upsilon_{1}^{2}  \nonumber \\
 & & -\epsilon_{1}\epsilon_{3}\frac{\upsilon_{\tilde{\nu}_{e}}}{\upsilon_{\tilde{\nu}_{\tau}}} - 
\epsilon_{2}\epsilon_{3}\frac{\upsilon_{\tilde{\nu}_{\mu}}}{\upsilon_{\tilde{\nu}_{\tau}}} - 
\frac{g^{2}}{4}(\upsilon_{\tilde{\nu}_{e}}^{2}+\upsilon_{\tilde{\nu}_{\mu}}^{2}), \nonumber \\
{\cal M}_{c 5,6}^{2} &=& 0 ,\nonumber \\
{\cal M}_{c 5,7}^{2} &=& 0 ,\nonumber \\
{\cal M}_{c 5,8}^{2} &=& \frac{1}{\sqrt{2}}l_{3}\mu\upsilon_{\tilde{\nu}_{\tau}} 
- \frac{1}{\sqrt{2}}l_{s_{3}}\mu\upsilon_{1} ,\nonumber \\
{\cal M}_{c 6,6}^{2} &=& -\frac{g'^{2}}{4}(\upsilon_{1}^{2}-\upsilon_{2}^{2}+\sum_{I}\upsilon
_{\tilde{\nu}_{I}}^{2})+\frac{1}{2}l_{1}^{2}(\upsilon_{1}^{2} + \upsilon_{\tilde{\nu}_{e}}^{2})
+ m_{R^{1}}^{2} ,\nonumber  \\
{\cal M}_{c 6,7}^{2} &=& \frac{1}{2}l_{1}l_{2}\upsilon_{\tilde{\nu}_{e}}\upsilon_{\tilde{\nu}_{\mu}} ,\nonumber \\
{\cal M}_{c 6,8}^{2} &=& \frac{1}{2}l_{1}l_{3}\upsilon_{\tilde{\nu}_{e}}\upsilon_{\tilde{\nu}_{\tau}} ,\nonumber \\
{\cal M}_{c 7,7}^{2} &=& -\frac{g'^{2}}{4}(\upsilon_{1}^{2}-\upsilon_{2}^{2}+\sum_{I}\upsilon
_{\tilde{\nu}_{I}}^{2})+\frac{1}{2}l_{1}^{2}(\upsilon_{1}^{2} + \upsilon_{\tilde{\nu}_{\mu}}^{2})
+ m_{R^{2}}^{2}   ,\nonumber  \\
{\cal M}_{c 7,8}^{2} &=& \frac{1}{2}l_{2}l_{3}\upsilon_{\tilde{\nu}_{\mu}}\upsilon_{\tilde{\nu}_{\tau}} ,\nonumber \\
{\cal M}_{c 8,8}^{2} &=& -\frac{g'^{2}}{4}(\upsilon_{1}^{2}-\upsilon_{2}^{2}+\sum_{I}\upsilon
_{\tilde{\nu}_{I}}^{2})+\frac{1}{2}l_{3}^{2}(\upsilon_{1}^{2} + \upsilon_{\tilde{\nu}_{\tau}}^{2})
+ m_{R^{3}}^{2}.  
\label{eq-21}
\end{eqnarray}
Note here that to obtain Eq.\ (\ref{eq-21}), Eq.\ (\ref{masspara}) is used sometimes.

\section{The mixing of the squarks}

In a general case, the matrix of the squarks mixing should be 6$\times$6. Under our assumptions, we do not consider the
squarks mixing between different generations. From superpotential Eq.\ (\ref{eq-2}) and the soft-breaking terms, we find
the up squarks mass matrix of the  I-th generation can be written as:
\begin{equation}
{\cal M}_{U^{I}}^{2} = 
\left(  
\begin{array}{cc}
\frac{1}{24}(3g^{2} - g^{\prime^{2}})(\upsilon^{2} - 2\upsilon_{2}^{2}) +
 \frac{u_{I}^{2}}{2}\upsilon_{2}^{2} + m_{Q^{I}}^{2} & 
\frac{1}{\sqrt{2}}(u_{I}\mu \upsilon_{1} - 
u_{I}\sum\limits_{J=1}^{3}\epsilon_{J}\upsilon_{\tilde{\nu}_{J}} - u_{S_{I}}\mu\upsilon_{2})  \\
\frac{1}{\sqrt{2}}(u_{I}\mu \upsilon_{1} - 
u_{I}\sum\limits_{J=1}^{3}\epsilon_{J}\upsilon_{\tilde{\nu}_{J}} - u_{S_{I}}\mu\upsilon_{2})  &
\frac{1}{6}g^{\prime^{2}}(\upsilon^{2} - 2\upsilon_{2}^{2}) + 
\frac{u_{I}^{2}}{2}\upsilon_{2}^{2} + m_{U^{I}}^{2}
\end{array} \right)
\label{matrix-upsqu}
\end{equation}
where $I=(1$, $2$, $3)$ is the index of the generations. 
The current eigenstates $\tilde{Q}_{1}^{I}$ and $\tilde{U}^{I}$ connect to the two physical (mass) 
eigenstates $\tilde{U}_{I}^{i}$$(i=(1$, $2)$ through
\begin{equation}
\tilde{U}_{I}^{i} = Z_{U^{I}}^{i,1}\tilde{Q}_{1}^{I} + Z_{U^{I}}^{i,2}\tilde{U}^{I}
\label{upsqu-mix}
\end{equation}
and $Z_{U^{I}}$ is determined by the condition:
\begin{equation}
Z_{U^{I}}^{\dag}{\cal M}_{U^{I}}^{2}Z_{U^{I}} = {\rm diag}(M_{U_{I}^{1}}^{2}, M_{U_{I}^{2}}^{2})
\end{equation}
In a similar way, we can give the down squarks mass matrix of the  I-th generation:
\begin{equation}
{\cal M}_{D^{I}}^{2} =
\left(  
\begin{array}{cc}
-\frac{1}{24}(3g^{2} + g^{\prime^{2}})(\upsilon^{2} - 2\upsilon_{2}^{2}) + 
\frac{d_{I}^{2}}{2}\upsilon_{1}^{2} + m_{Q^{I}}^{2} & 
-\frac{1}{\sqrt{2}}(d_{I}\mu \upsilon_{2} - d_{S_{I}}\mu\upsilon_{1})  \\
-\frac{1}{\sqrt{2}}(d_{I}\mu \upsilon_{2} - d_{S_{I}}\mu\upsilon_{1})  &
-\frac{1}{12}g^{\prime^{2}}(\upsilon^{2} - 2\upsilon_{2}^{2}) + \frac{d_{I}^{2}}{2}\upsilon_{1}^{2} + m_{D^{I}}^{2}
\end{array} \right)
\label{matrix-downsqu}
\end{equation}
The fields $\tilde{Q}_{2}^{I}$ and $\tilde{D}^{I}$ relate to the two physical (mass)
eigenstates $\tilde{D}_{I}^{i}$ $(i=(1$, $2)$:
\begin{eqnarray}
&&\tilde{D}_{I}^{i} = Z_{D^{I}}^{i,1}\tilde{Q}_{2}^{I} + Z_{D^{I}}^{i,2}\tilde{D}^{I}  \nonumber \\
&&Z_{D^{I}}^{\dag}{\cal M}_{D^{I}}^{2}Z_{D^{I}} = {\rm diag}(M_{D_{I}^{1}}^{2}, M_{D_{I}^{2}}^{2})
\end{eqnarray}

\section{Expressions of the couplings in ${\cal L}_{SSS}$ and ${\cal L}_{SSSS}$}

In this appendix, we give precise expressions of the couplings that 
appear in the ${\cal L}_{SSS}$ and ${\cal L}_{SSSS}$.
The method has been described clearly in text, the results are:
\begin{eqnarray}
A_{ec}^{kij} &=& \frac{g^{2} + g^{\prime^{2}}}{4}\bigg(\upsilon_{1}Z_{even}^{k,1}Z_{c}^{i,1}Z_{c}^{j,1} + 
\upsilon_{2}Z_{even}^{k,2}Z_{c}^{i,2}Z_{c}^{j,2} + 
\sum\limits_{I=1}^{3}\upsilon_{\tilde{\nu}_{I}}Z_{even}^{k,2+I}Z_{c}^{i,2+I}
Z_{c}^{j,2+I}\bigg)  \nonumber \\
 & & +\sum\limits_{I=1}^{3}\bigg(\frac{
g^{\prime^{2}} - g^{2}}{4} + l_{I}^{2}\bigg)\bigg(\upsilon_{1}Z_{even}^{k,1}Z_{c}^{i,2+I}Z_{c}^{j,2+I} 
+ \upsilon_{\tilde{\nu}_{I}}Z_{even}^{k,2+I}
Z_{c}^{i,1}Z_{c}^{j,1}\bigg)  \nonumber \\
 & &+\sum\limits_{I=1}^{3}\bigg(\frac{g^{2}}{4} - \frac{1}{2}l_{I}^{2}\bigg)\bigg\{
\upsilon_{1}Z_{even}^{k,2+I}\Big(Z_{c}^{i,2+I}Z_{c}^{j,2} + 
Z_{c}^{i,2}Z_{c}^{j,2+I}\Big) + \upsilon_{\tilde{\nu}_{I}}Z_{even}^{k,1}\Big(Z_{c}^{i,2+I}Z_{c}^{j,2}  \nonumber \\
& & + Z_{c}^{i,2}Z_{c}^{j,2+I}\Big)\bigg\} + \frac{g^{2} - g^{\prime^{2}}}{4}\bigg(
\upsilon_{1}Z_{even}^{k,1}Z_{c}^{i,2}Z_{c}^{j,2} + \upsilon_{2}Z_{even}^{k,2}Z_{c}^{i,1}Z_{c}^{j,1}  \nonumber \\
 & & +\sum\limits_{I=1}^{3}\upsilon_{\tilde{\nu}_{I}}Z_{even}^{k,2+I}Z_{c}^{i,2}Z_{c}^{j,2} +
\upsilon_{2}Z_{even}^{k,2}Z_{c}^{i,2+I}Z_{c}^{j,2+I}\bigg)  \nonumber \\
 & & +\sum\limits_{I=1}^{3}\bigg[\bigg(l_{I}^{2} - \frac{g^{\prime^{2}}}{2}\bigg)
\upsilon_{\tilde{\nu}_{I}}Z_{even}^{k,2+I}
Z_{c}^{i,5+I}Z_{c}^{j,5+I} + \bigg(l_{I}^{2} - \frac{g^{\prime^{2}}}{2}\bigg)
\upsilon_{1}Z_{even}^{k,1}Z_{c}^{i,5+I}Z_{c}^{j,5+I}  \nonumber \\
 & &+\frac{g^{\prime^{2}}}{2}\upsilon_{2}Z_{even}^{k,2}Z_{c}^{i,5+I}Z_{c}^{j,5+I}\bigg] +  
\frac{g^{2}}{4}\bigg(\upsilon_{\tilde{\nu}_{I}}Z_{even}^{k,2} + 
\upsilon_{2}Z_{even}^{k,2+I}\bigg)\bigg(Z_{c}^{i,2+I}Z_{c}^{j,2}  \nonumber \\
 & & +Z_{c}^{i,2}Z_{c}^{j,2+I}\bigg) + \frac{g^{2}}{4}\bigg(\upsilon_{1}Z_{even}^{k,2} + 
\upsilon_{2}Z_{even}^{k,1}\bigg)\bigg(Z_{c}^{i,1}Z_{c}^{j,2}  \nonumber \\
 & & +Z_{c}^{i,2}Z_{c}^{j,1}\bigg) + \frac{1}{\sqrt{2}}l_{I}
\epsilon_{3}Z_{even}^{k,1}\bigg(Z_{c}^{i,4}Z_{c}^{j,2}   \nonumber \\
 & & +Z_{c}^{i,2}Z_{c}^{j,4} \bigg) + \frac{1}{\sqrt{2}}
\sum\limits_{I=1}^{3}l_{I}\epsilon_{I}Z_{even}^{k,2}\bigg(
Z_{c}^{i,5+I}Z_{c}^{j,1} + Z_{c}^{i,1}Z_{c}^{j,5+I} \bigg)  \nonumber \\
A_{oc}^{kij} &=& \sum\limits_{I=1}^{3}\Bigg\{\bigg(\frac{g^{2}}{4} - l_{I}^{2}\bigg)
\bigg[\upsilon_{\tilde{\nu}_{I}}Z_{odd}^{k,1}\Big(Z_{c}^{i,1}
Z_{c}^{j,2+I} - Z_{c}^{i,2+I}Z_{c}^{j,1}\Big) \nonumber \\
&& + \upsilon_{1}Z_{odd}^{k,2+I}\Big(Z_{c}^{i,1}Z_{c}^{j,2+I} - 
Z_{c}^{i,2+I}Z_{c}^{j,1}\Big)\bigg]  \nonumber \\
 & & + \frac{g^{2}}{4}\Big(\upsilon_{\tilde{\nu}_{I}}Z_{odd}^{k,2} + 
\upsilon_{2}Z_{odd}^{k,2+I}\Big)\Big(Z_{c}^{i,2+I}Z_{c}^{j,2} - 
Z_{c}^{i,2}Z_{c}^{j,2+I}\Big)
+ \frac{g^{2}}{4}\Big(\upsilon_{\tilde{\nu}_{I}}Z_{odd}^{k,2}  \nonumber \\
 & & +\upsilon_{2}Z_{odd}^{k,2+I}\Big)\Big(Z_{c}^{i,2+I}Z_{c}^{j,2} - 
Z_{c}^{i,2}Z_{c}^{j,2+I}\Big)  \nonumber \\
 & & \frac{1}{\sqrt{2}}l_{I}\epsilon_{I}Z_{odd}^{k,1}\Big(- Z_{c}^{i,5+I}Z_{c}^{j,2} 
+ Z_{c}^{i,2}Z_{c}^{j,5+I}\Big)  \nonumber \\
 & & -\frac{1}{\sqrt{2}}l_{I}\epsilon_{I}Z_{odd}^{k,2}\Big( Z_{c}^{i,5+I}Z_{c}^{j,1} 
- Z_{c}^{i,1}Z_{c}^{j,5+I}\Big) \Bigg\} \nonumber  \\
{\cal A}_{ec}^{klij} &=& \frac{g^{2} + g^{\prime^{2}}}{8}\bigg(Z_{even}^{k,1}Z_{even}^{l,1}Z_{c}^{i,1}Z_{c}^{j,1} + 
Z_{even}^{k,2}Z_{even}^{l,2}Z_{c}^{i,2}Z_{c}^{j,2} + 
\sum\limits_{I=1}^{3}Z_{even}^{k,2+I}Z_{even}^{l,2+I}Z_{c}^{i,2+I}Z_{c}^{j,2+I}\bigg) \nonumber \\
 & &+\sum\limits_{I=1}^{3}\bigg(\frac{g^{\prime^{2}} - g^{2}}{8} + \frac{1}{2}l_{I}^{2}\bigg)
\bigg(Z_{even}^{k,1}Z_{even}^{l,1}Z_{c}^{i,2+I}Z_{c}^{j,2+I} +
Z_{even}^{k,2+I}Z_{even}^{l,2+I}Z_{c}^{i,1}Z_{c}^{j,1}\bigg)  \nonumber \\
 & &+\frac{g^{\prime^{2}} - g^{2}}{8}\bigg[Z_{even}^{k,1}Z_{even}^{l,1}Z_{c}^{i,2}Z_{c}^{j,2} 
+ Z_{even}^{k,2}Z_{even}^{l,2}Z_{c}^{i,1}Z_{c}^{j,1}  \nonumber \\
 & & +\sum\limits_{I=1}^{3}\Big(Z_{even}^{k,2+I}Z_{even}^{l,2+I}Z_{c}^{i,2}Z_{c}^{j,2} 
+  Z_{even}^{k,2}Z_{even}^{l,2}
Z_{c}^{i,2+I}Z_{c}^{j,2+I}\Big)\bigg]  \nonumber  \\
 & & +\frac{g^{2}}{4}\bigg[Z_{even}^{k,1}Z_{even}^{l,2}\Big(Z_{c}^{i,2}Z_{c}^{j,1} + Z_{c}^{i,1}Z_{c}^{j,2}) + 
\sum\limits_{I=1}^{3}Z_{even}^{k,1}Z_{even}^{l,2+I}(Z_{c}^{i,2+I}Z_{c}^{j,1}  \nonumber \\
 & &+Z_{c}^{i,1}Z_{c}^{j,2+I}\Big) + \sum\limits_{I=1}^{3}Z_{even}^{k,2}Z_{even}^{l,2+I}
\Big(Z_{c}^{i,2+I}Z_{c}^{j,2} + 
Z_{c}^{i,2}Z_{c}^{j,2+I}\Big) \bigg]  \nonumber  \\
 & & -\sum\limits_{I=1}^{3}\bigg[\frac{l_{I}^{2}}{2}Z_{even}^{k,1}Z_{even}^{l,2+I}\Big(Z_{c}^{i,2+I}Z_{c}^{j,1} + 
Z_{c}^{i,1}Z_{c}^{j,2+I}\Big) - \frac{g^{\prime^{2}}}{4}\Big(
Z_{even}^{k,1}Z_{even}^{l,1}Z_{c}^{i,5+I}Z_{c}^{j,5+I}  \nonumber \\
 & &- Z_{even}^{k,2}Z_{even}^{l,2}Z_{c}^{i,5+I}Z_{c}^{j,5+I}
+ Z_{even}^{k,2+I}Z_{even}^{l,2+I}Z_{c}^{i,5+I}Z_{c}^{j,5+I}\Big)  \nonumber  \\
 & & +\frac{l_{I}^{2}}{2}Z_{even}^{k,1}Z_{even}^{l,1}Z_{c}^{i,5+I}Z_{c}^{j,5+I} \bigg]  \nonumber  \\
{\cal A}_{oc}^{klij} &=& \frac{g^{2} + g^{\prime^{2}}}{8}\Big(Z_{odd}^{k,1}Z_{odd}^{l,1}Z_{c}^{i,1}Z_{c}^{j,1} + 
Z_{odd}^{k,2}Z_{odd}^{l,2}Z_{c}^{i,2}Z_{c}^{j,2} \nonumber \\
 & &+\sum\limits_{I=1}^{3}Z_{odd}^{k,2+I}Z_{odd}^{l,2+I}Z_{c}^{i,2+I}Z_{c}^{j,2+I}\Big) + 
\sum\limits_{I=1}^{3}\bigg(\frac{g^{\prime^{2}} - g^{2}}{8} + \frac{1}{2}l_{I}^{2}\bigg)
\bigg(Z_{odd}^{k,1}Z_{odd}^{l,1}Z_{c}^{i,2+I}Z_{c}^{j,2+I}  \nonumber \\
 & &+Z_{odd}^{k,2+I}Z_{odd}^{l,2+I}Z_{c}^{i,1}Z_{c}^{j,1}\bigg) + 
\frac{g^{\prime^{2}} - g^{2}}{8}\bigg\{Z_{odd}^{k,1}Z_{odd}^{l,1}Z_{c}^{i,2}Z_{c}^{j,2}  \nonumber \\
 & &+Z_{odd}^{k,2}Z_{odd}^{l,2}
Z_{c}^{i,1}Z_{c}^{j,1} + \sum\limits_{I=1}^{3}\Big(Z_{odd}^{k,2+I}Z_{odd}^{l,2+I}Z_{c}^{i,2}Z_{c}^{j,2}  \nonumber \\
 & & +Z_{odd}^{k,2}Z_{odd}^{l,2}Z_{c}^{i,2+I}Z_{c}^{j,2+I}\Big)\bigg\} +
\frac{g^{2}}{4}\Bigg\{Z_{odd}^{k,1}Z_{odd}^{l,2}\Big(Z_{c}^{i,2}Z_{c}^{j,1} + Z_{c}^{i,1}Z_{c}^{j,2}\Big)  \nonumber \\
 & &+ \sum\limits_{I=1}^{3}\bigg[Z_{odd}^{k,1}Z_{odd}^{l,3}\Big(Z_{c}^{i,3}Z_{c}^{j,1} + Z_{c}^{i,1}Z_{c}^{j,3}\Big) + 
Z_{odd}^{k,2}Z_{odd}^{l,3}\Big(Z_{c}^{i,3}Z_{c}^{j,2} + Z_{c}^{i,2}Z_{c}^{j,3}\Big)\bigg] \Bigg\} \nonumber  \\
 & & -\sum\limits_{I=1}^{3}\Bigg\{\frac{l_{I}^{2}}{2}Z_{odd}^{k,1}Z_{odd}^{l,2+I}
\Big(Z_{c}^{i,2+I}Z_{c}^{j,1} + Z_{c}^{i,1}Z_{c}^{j,2+I}\Big) - 
\frac{g^{\prime^{2}}}{4}\Big(Z_{odd}^{k,1}Z_{odd}^{l,1}Z_{c}^{i,5+I}Z_{c}^{j,5+I}  \nonumber \\
 & & -Z_{odd}^{k,2}Z_{odd}^{l,2}Z_{c}^{i,5+I}Z_{c}^{j,5+I}
+ Z_{odd}^{k,2+I}Z_{odd}^{l,2+I}Z_{c}^{i,5+I}Z_{c}^{j,5+I}\Big)  \nonumber  \\
 & & +\frac{l_{I}^{2}}{2}Z_{odd}^{k,1}Z_{odd}^{l,1}Z_{c}^{i,5+I}Z_{c}^{j,5+I}\Bigg\}  \nonumber  \\
{\cal A}_{eoc}^{klij} &=& \sum\limits_{I=1}^{3}\bigg[\Big(\frac{g^{2}}{8} + 
\frac{1}{2}l_{I}^{2}\Big)\Big(Z_{even}^{k,2+I}Z_{odd}^{l,1} + Z_{even}^{k,1}Z_{odd}^{l,2+I}\Big)
\Big(Z_{c}^{i,1}Z_{c}^{j,2+I} - Z_{c}^{i,2+I}Z_{c}^{j,1}\Big) \nonumber \\
 & &-\frac{g^{2}}{8}\Big(Z_{even}^{k,2}Z_{odd}^{l,2+I} + Z_{even}^{k,2+I}Z_{odd}^{l,2}\Big)
\Big(Z_{c}^{i,2+I}Z_{c}^{j,2} - Z_{c}^{i,2}Z_{c}^{j,2+I}\Big)\bigg]  \nonumber \\
 & & -\frac{g^{2}}{8}\Big(Z_{even}^{k,1}Z_{odd}^{l,2} + Z_{even}^{k,2}Z_{odd}^{l,1}\Big)
\Big(Z_{c}^{i,1}Z_{c}^{j,2} - 
Z_{c}^{i,2}Z_{c}^{j,2}\Big)  \nonumber \\
{\cal A}_{cc}^{ijkl} &=&\frac{g^{2} + g^{\prime^{2}}}{8}\bigg[
\sum\limits_{m,n=1}^{5}Z_{c}^{i,m}Z_{c}^{j,m}Z_{c}^{k,n}Z_{c}^{l,n} + 
2\sum\limits_{I=1}^{3}Z_{c}^{i,2+I}Z_{c}^{j,2+I}\Big(Z_{c}^{k,1}Z_{c}^{l,1} - Z_{c}^{k,2}Z_{c}^{l,2}\Big) \nonumber \\
 & & -2Z_{c}^{i,1}Z_{c}^{j,1}Z_{c}^{k,2}Z_{c}^{l,2} \bigg] +
\frac{g^{\prime^{2}}}{2}\bigg[\sum\limits_{I=1}^{3}Z_{c}^{i,5+I}Z_{c}^{j,5+I}\Big(
 - Z_{c}^{k,5+I}Z_{c}^{l,5+I} - Z_{c}^{k,2+I}Z_{c}^{l,2+I}\Big)  \nonumber \\
 & & -\sum\limits_{I=1}^{3}Z_{c}^{i,5+I}Z_{c}^{j,5+I}Z_{c}^{k,1}Z_{c}^{l,1}\bigg] + 
\sum\limits_{I=1}^{3}l_{I}^{2}Z_{c}^{i,5+I}Z_{c}^{j,5+I}Z_{c}^{k,1}Z_{c}^{l,1} \nonumber 
\end{eqnarray}
where the mixing matrices $Z_{even}$, $Z_{odd}$ and $Z_{c}$ are defined as in Eq.\ (\ref{masseven}), Eq.\ (\ref{oddhiggs}) 
and Eq.\ (\ref{charhiggs}).

\begin{figure}
\setlength{\unitlength}{1mm}
\begin{picture}(140,180)(30,30)
\put(-10,-10){\includegraphics{fig.3}}
\end{picture}
\caption[]{Feynman rules for SSVV vertices. Part(I)}
\label{fig3}
\end{figure}

\begin{figure}
\setlength{\unitlength}{1mm}
\begin{picture}(140,180)(30,30)
\put(-10,-10){\includegraphics{fig.4}}
\end{picture}
\caption[]{Feynman rules for SSVV vertices. Part(II)}
\label{fig4}
\end{figure}

\begin{figure}
\setlength{\unitlength}{1mm}
\begin{picture}(140,180)(30,30)
\put(-10,-10){\includegraphics{fig.5}}
\end{picture}
\caption[]{Feynman rules for the self-coupling of Higgs. Part(I)}
\label{fig5}
\end{figure}

\begin{figure}
\setlength{\unitlength}{1mm}
\begin{picture}(140,180)(30,30)
\put(-10,-10){\includegraphics{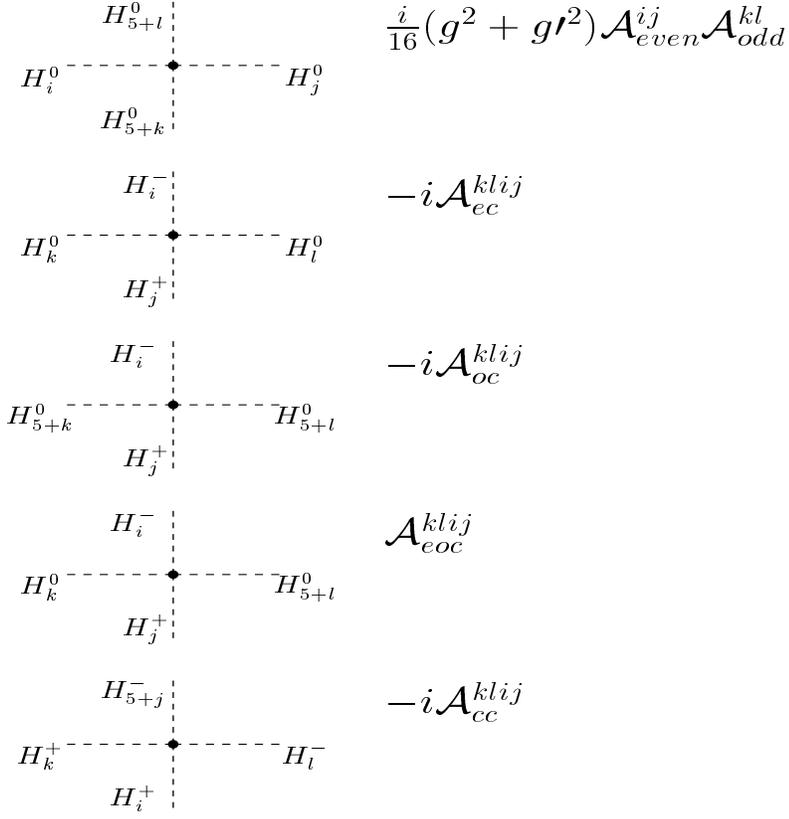}}
\end{picture}
\caption[]{Feynman rules for the self-coupling of Higgs. Part(II)}
\label{fig6}
\end{figure}

\begin{figure}
\setlength{\unitlength}{1mm}
\begin{picture}(140,180)(30,30)
\put(-10,-10){\includegraphics{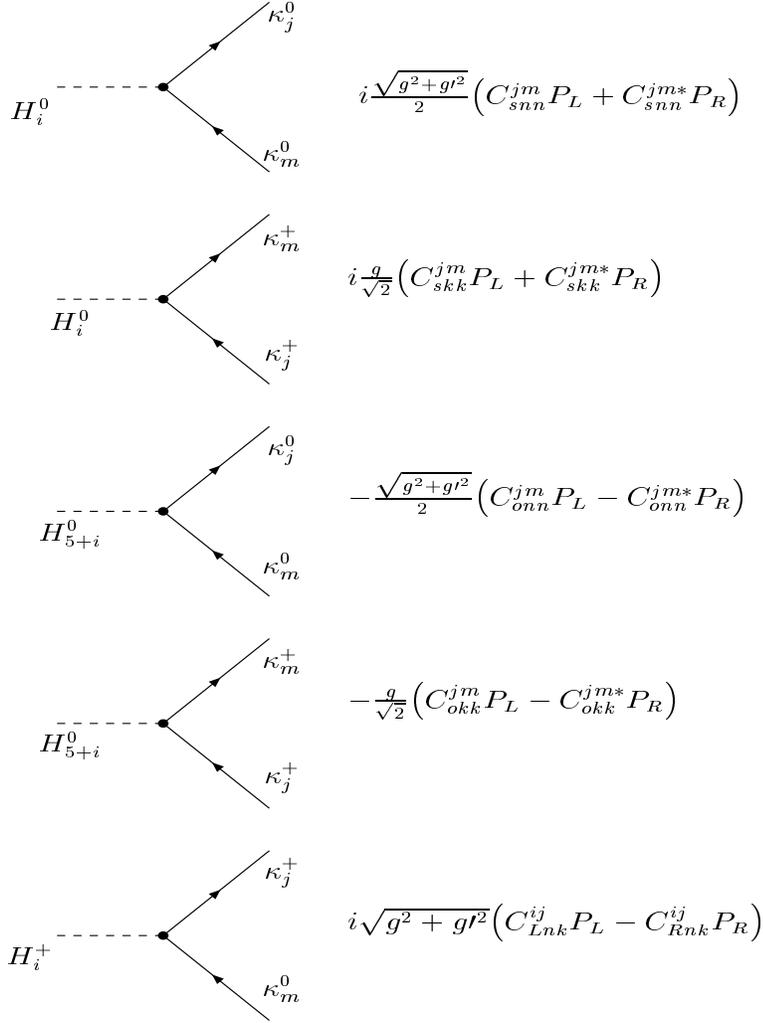}}
\end{picture}
\caption[]{Feynman rules for the coupling of Higgs with charginos or neutralinos.}
\label{fig7}
\end{figure}

\begin{figure}
\setlength{\unitlength}{1mm}
\begin{picture}(140,180)(30,30)
\put(-10,-10){\includegraphics{fig.8}}
\end{picture}
\caption[]{Feynman rules for the coupling of gauge bosons with charginos or neutralinos.}
\label{fig8}
\end{figure}

\begin{figure}
\setlength{\unitlength}{1mm}
\begin{picture}(140,180)(30,30)
\put(-10,-10){\includegraphics{fig.9}}
\end{picture}
\caption[]{Feynman rules for the coupling of quarks, squarks with charginos or neutralinos.}
\label{fig9}
\end{figure}

\begin{figure}
\setlength{\unitlength}{1mm}
\begin{picture}(140,180)(30,30)
\put(-10,20){\includegraphics{fig.10}}
\end{picture}
\caption[]{The mass of the lightest neutralino varied with X. The parameters are assigned as $m_{1}=m_{2}=3000$GeV
and (a)$\tan\beta = 20$, $\tan\theta_{\upsilon} = 20$; (b)$\tan\beta = 20$, $\tan\theta_{\upsilon} = 0.5$; 
(c)$\tan\beta = 0.5$, $\tan\theta_{\upsilon} = 20$; (d)$\tan\beta = 0.5$, $\tan\theta_{\upsilon} = 0.5$. The dot-dash lines
correspond to $m_{\nu_{\tau}}=0.2$MeV, the dash lines correspond to $m_{\nu_{\tau}}=2$MeV, and dot lines correspond
to $m_{\nu_{\tau}}=20$MeV}
\label{fig10}
\end{figure}

\begin{figure}
\setlength{\unitlength}{1mm}
\begin{picture}(140,180)(30,30)
\put(-10,20){\includegraphics{fig.11}}
\end{picture}
\caption[]{The mass of the lightest chargino varied with X. The parameters are assigned as $m_{1}=m_{2}=3000$GeV
and (a)$\tan\beta = 20$, $\tan\theta_{\upsilon} = 20$; (b)$\tan\beta = 20$, $\tan\theta_{\upsilon} = 0.5$; 
(c)$\tan\beta = 0.5$, $\tan\theta_{\upsilon} = 20$; (d)$\tan\beta = 0.5$, $\tan\theta_{\upsilon} = 0.5$. The dot-dash lines
correspond to $m_{\nu_{\tau}}=0.2$MeV, the dash lines correspond to $m_{\nu_{\tau}}=2$MeV, and dot lines correspond
to $m_{\nu_{\tau}}=20$MeV}
\label{fig11}
\end{figure}

\begin{figure}
\setlength{\unitlength}{1mm}
\begin{picture}(140,180)(30,30)
\put(-10,20){\includegraphics{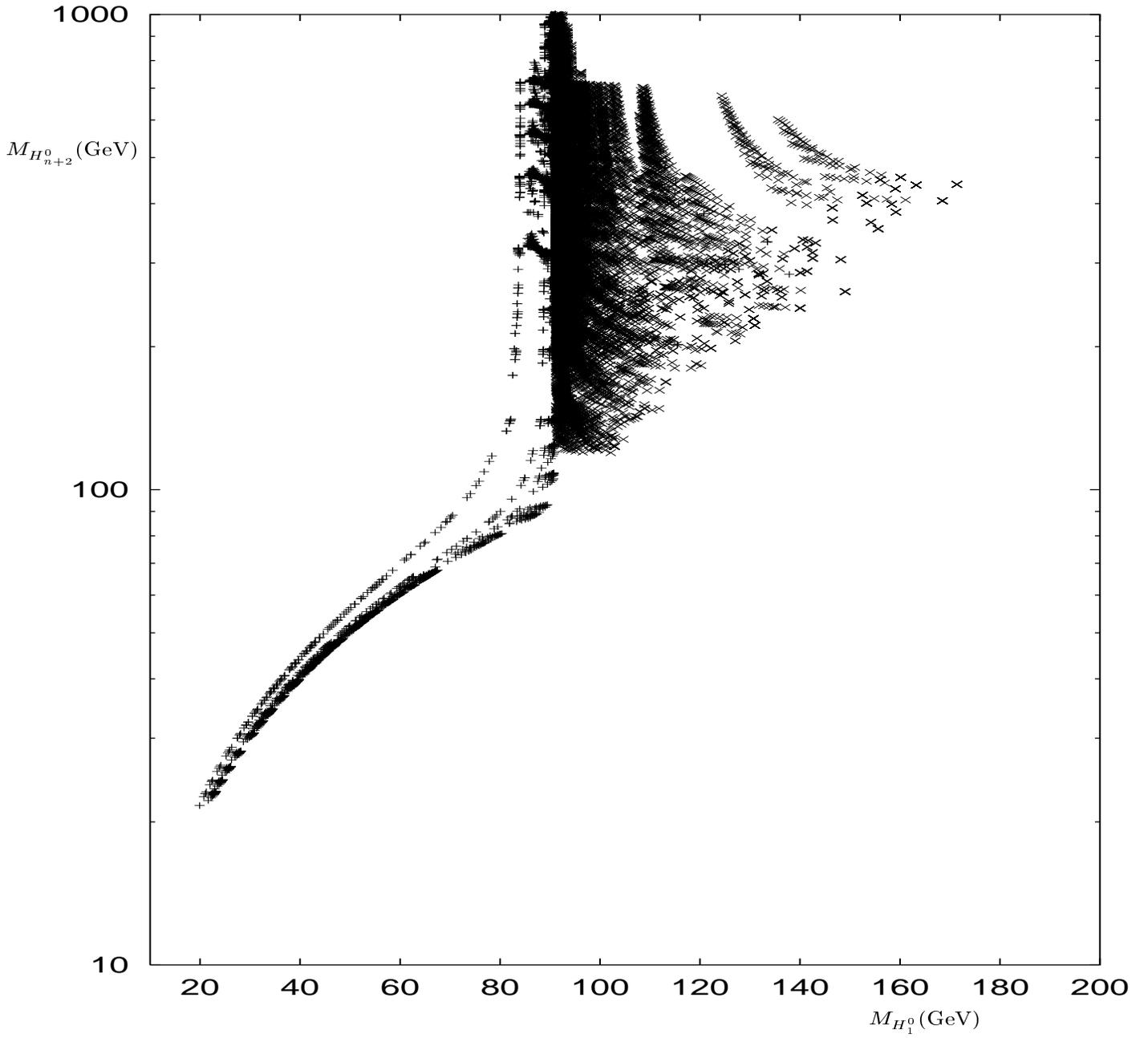}}
\end{picture}
\caption[]{The mass of the lightest CP-odd Higgs varied with the mass of the lightest CP-even Higgs($n=3$). 
The range of parameters are assigned as -$10^{5}$GeV$^{2}$ $\leq $ $X_{s}$, $Y_{s}$, $Z_{s}$ $\leq $ $10^{5}$ GeV$^{2}$
and $0.5 \leq \tan\beta \leq 50$, $0.5 \leq \tan\theta_{\upsilon} \leq 50$.}
\label{fig12}
\end{figure}

\end{document}